\documentclass[aa]{aa}
\bibpunct{(}{)}{;}{a}{}{,} 
\usepackage{graphicx}
\usepackage{comment}
\usepackage{txfonts}
\usepackage{stfloats}
\usepackage{algorithm}
\usepackage{algpseudocode}

\def\apgt{\ {\raise-.5ex\hbox{$\buildrel>\over\sim$}}\ }
\def\aplt{\ {\raise-.5ex\hbox{$\buildrel<\over\sim$}}\ }
\def\lteq{\ {\raise-.5ex\hbox{$\buildrel<\over-$}}\ }

\bibpunct{(}{)}{;}{a}{}{,} 

\def\aap{\ {A\&A}\ }

\def\apjl{\ {ApJL}\ }
\def\apjs{\ {ApJS}\ }

\def\arxivprefixe{arXiv}
\def\arxivprefixesep{:}
\algnewcommand{\LineComment}[1]{\State \(\triangleright\) #1}

\begin{document} 

\title{{\sc Venice}: a multi-scale operator-splitting algorithm for multi-physics simulations}
          \author{
          Maite J. C. Wilhelm\inst{1}
          \and
          S. Portegies Zwart\inst{1}
          }
   \institute{
             Leiden Observatory, University of Leiden, 
             Niels Bohrweg 2, 2333 CA Leiden, the Netherlands\\
             \email{wilhelm@strw.leidenuniv.nl}
             \email{spz@strw.leidenuniv.nl}
             }
   \date{Received XXXXX; accepted XXXXXX}


   \abstract
  {We present {\sc Venice}, an operator splitting algorithm to
    integrate a numerical model on a hierarchy of timescales.}
  {{\sc Venice} allows a wide variety of different physical processes
    operating a different scales to be coupled on individual and
    adaptive time-steps. It therewith mediates the development of
    complex multi-scale and multi-physics simulation environments with
    a wide variety of independent components.}
  {The coupling between various physical models and scales is dynamic,
    and realized through (Strang) operators splitting using adaptive
    time steps.  }
  {We demonstrate the functionality and performance of
    this algorithm using astrophysical models of a stellar cluster,
    first coupling gravitational dynamics and stellar evolution, then
    coupling internal gravitational dynamics with dynamics within a
    galactic background potential, and finally combining these models
    while also introducing dwarf galaxy-like perturbers. These tests
    show numerical convergence for decreasing coupling timescales,
    demonstrate how {\sc Venice} can improve the performance of a
    simulation by shortening coupling timescales when appropriate, and
    provide a case study of how {\sc Venice} can be used to gradually
    build up and tune a complex multi-physics model. Although the
    examples couple complete numerical models, {\sc Venice} can also
    be used to efficiently solve systems of stiff differential
    equations.  }
  {}

            
\maketitle

\section{Introduction}

Improvements in computational capabilities over recent decades have
allowed numerical models to increase in scale and complexity. This has
mostly come in the form of models covering a range of physical
processes -- multi-physics models -- and physical scales multi-scale
models.

Many multi-physics models use the concept of operator
splitting. Instead of modelling all physical processes in concert,
they are solved sequentially.  Operator splitting introduces a
numerical error, but this error typically decreases as the timescale
on which the operators are split decreases. If the operators are
applied sequentially for the same time-steps, the error is linear in
the time-steps. Applying the operators for fractional time steps in
interwoven patterns, much like a Verlet scheme \cite{PhysRev.159.98},
results in errors quadratic in the time step \citep[e.g Strang
  splitting,][]{Strang1968}. Higher-order operator splitting methods
exist, but often in limited cases or non-trivial forms
\citep{Jia2011,Christlieb2015,PortegiesZwart2020}.

Even with traditional Strang splitting, all the operators are applied
at roughly the same timescale, which necessarily must be the shortest
common timescale of the system. In addition, the adopted time step
between two solvers is symmetric.  As multi-physics models cover
increasingly more physical processes, they tend to become
increasingly more multi-scale as well. In addition, the range of
scales in the problem may dramatically change with time.  In that
case, a time step that dynamically varies with time and across
coupling method leads to considerable speed-up and smaller
errors.

This issue of a large range of timescales has long been recognised in
modelling systems with a large number of self-gravitating bodies
\citep{1985mts..conf..377A}. In response, schemes were developed where
particles could be evolved on their individual, appropriate,
time-steps
\citep{Wielen1967,McMillan1986,Pelupessy2012a}. Additionally, the {\sc
  bridge} method \citep{Fujii2007} has been developed to couple
gravitationally interacting systems of different physical scales.  It
splits the combined (Hamiltonian) operator into two isolated,
self-consistent systems and interaction terms. The two systems can be
solved independently, using the preferred methods for each, but they
interact on larger timescales.  Although quite powerful, {\sc bridge},
has several important shortcomings.  These include its fixes and
constant coupling time scale, the rigidity of coupled schemes used,
and the fixation of the data set on which each differential solver
operates. As a consequence, the {\sc bridge} once constructed, is
rigid in the spatial and temporal domain, and in methodology. This
rigidity limits its application in modern simulation endeavors.

With the gradual increase in the sophistication of domain-specific
numerical solvers, it has become increasingly challenging to develop
simulation models incorporating new physics, and employing new
algorithms. Expanding those model also has become more complex,
because the higher sophistication renders the underlying numerical
framework more fragile. Without thorough modularization, we move into
a direction in which simulation codes become unwieldingly complex.

In this paper we introduce {\sc Venice}, an algorithm to split an
arbitrary number of operators on a wide range of timescales. We
explore coupling of operators on dynamical variable timescales, and
discuss how {\sc bridge} is incorporated into the framework.  We
present a reference implementation in the {\sc amuse} framework, using
it to demonstrate the capabilities and measure the performance of the
algorithm. {\sc Venice} improves performance in multi-physics
simulations by allowing different components to couple at individually
relevant timescales instead of the shortest common timescale.
Possibly even more important than performance is the possibility to
maintain the framework, verify its working and validate the results.
A operational version including research project using {\sc Venice} is
presented in \cite{Shuo2etal2024}, where we develope a planet
population synthesis analysis of planet formation in young stellar
clusters. We start, however, with a brief overview of exiting
frameworks for addressing (astro)physical problems with their
advantages and disadvantages.

\section{An brief overview of multi-scale simulation frameworks}

\citet{PortegiesZwart2018} argued for a modular approach to
multi-scale, multi-physics simulations.  He envisions an environment
in which highly optimised modules address with a limited scope are
combined, using a common interface, into a more complex framework that
cover the full palette of physical processes at their appropriate
scale, spatially and temporally.  By incorporating at least two
physically motivated solvers per domain, such a framework would
subsequently complies to ``Noah's Ark'' philosophy for code coupling
paradigms \citep{PortegiesZwart2018}.  Having two or more solvers for
addressing the same physical domain at the same spatial or temporal
scale allows for independent verification and validation of the
specific part of the simulation.

While the whole purpose of performing numerical simulations is to go
beyond what we can test from fundamental principles, an ``Noah's Ark''
principle should be considered one of the fundamental aspects of a
simulation framework environment.

We will briefly discuss a number of such frameworks, which comply in
part to the above criteria.  A more extensive overview of such
frameworks across disciplines is presented in
\cite{2015MultiscaleframeworksB}.  Each of these frameworks has its
advantages and focus area.  For example, see
\cite{10.1177/0037549719881204} for simulating power plant reactors,
     {\sc DEM} for computational fluid dynamics
     \citep{2018arXiv180808028P}, aluminum electrolysis
     \citep{EINARSRUD20173}, or artery dynamics in blood flow
     \citep{Veen2020} (discussed below), and the general implicit
     coupling framework for multi-physics problems
     \citep{10.2118/182714-MS}.

We will briefly discuss the three frameworks that most closely comply
to the listed requirements for a multi-purpose simulation environment.
These include {\sc Cactus}, the multiscale modeling framework, and
{\sc amuse}.

{\sc Cactus} \citep{10.1007/3-540-36569-9_13} is a general purpose problem
solving environment for the physical sciences \citep{allen2001cactus}.
It is modular and supports parallel computing across the network with
a variety of hardware.  {\sc Cactus} had a last update (version 4.12 of May
2022, see \url{http://www.cactuscode.org}).  {\sc Cactus} offers several
advantages over other frameworks.  It uses a component-based
architecture, where the core functionality is provided by a central
"flesh" and various "thorns" (modules) that can be independently
developed, compiled, and loaded.  The thorns can be parallelized, for
multi-core architectures and distributed computing environments.  Its
high level of abstractions and interoperability with other tools
simplifies the development of complex simulations,

Closest to the {\sc amuse} philosophy is probably the Multiscale
Modeling and Simulation Framework
\citep[MMSF,][]{Borgdorff2013,Chopard2014}, which has been applied in
the {\sc muscle3} environment \citep{Veen2020}. However, in contrast
to {\sc amuse}'s structure of a driver script instructing a set of
models to run and exchange information, it instead manages a set of
distributed, mostly independent models which exchange messages via
externally configurable connections, which block if a result they need
is not yet available.

The Astrophysics Multiphysics Simulation Environment \cite[{\sc amuse}
  for short, ][]{PortegiesZwart2009,PortegiesZwart2013,Pelupessy2013}
has similar advantages as {\sc Cactus} and the {\sc MMSF}, it has the
additional advantage of dynamic adaptation and a high-level
programming interface. Although specifically designed for applications
in astronomy it branches out to oceanography and hydrology.  While
{\sc amuse} provides the capability of coupling an arbitrary number of
codes, it still relies on the coupling of an array of solvers with a
second-order scheme with constant time steps.  The {\sc amuse}
framework provides a first implementation of ``Noah's Ark'', providing
a pre-determined communication protocol with standardized rules and a
common data model.

{\sc amuse} adopts {\sc bridge} for its code coupling to first, second
and any even higher order \citep{PortegiesZwart2020}. But it fails on
time-dependent flexibility needed for modern multi-scale and
multi-physics simulations. Part of this problem has been addressed in
the {\sc nemesis} module in {\sc amuse} \citep{2019A&A...624A.120V}.

{\sc Nemesis} uses a global structure and sub-structures to couple the
stars and planetary system dynamics through a cascade of {\sc Bridge}
patterns. The global structure contains stars and planets integrated
by a direct $N$-body solver, while substructures are integrated using
a symplectic $N$-body solver coupled to the global system via a
symplectic {\sc Bridge} \cite{2020CNSNS..8505240P}. Nemesis dynamically
combines each planetary system's direct, symplectic and secular
integrators at runtime. Escaping planets are removed from the local
system and incorporated into the global $N$-body integrator, while the
remaining planets continue with their native solver.  This is a very
dynamic environment in which codes are spawned and killed at runtime.

With {\sc venice} we go further into ``Noah's Ark'' paradigm, by
allowing an operator based approach to code coupling. By relaxing the
coupling strategy, and allowing for a arbitrary number of codes to be
coupled with an arbitrary number of other codes with arbitrary
coupling times makes {\sc venice} more flexible, transparant and
modular than any of the above mentioned endeavors.

\section{Algorithm}

We here discuss the generalized algorithm for a multi-scale operator
approach.

\subsection{Operator splitting}

The {\sc Venice} algorithm is based on the concept of operator splitting. Consider a differential equation of the following form:

\begin{equation}
    \frac{dy}{dt} = L_1\left(y\right) + L_2\left(y\right).
\end{equation}

The exact solution to this differential equation is:

\begin{equation}
    y\left(\Delta t\right) = e^{\left(\hat{L}_1+\hat{L}_2\right)\Delta t}y\left(0\right).
\end{equation}

An approximate solution would be:

\begin{equation} \label{eq:linear}
    y\left(\Delta t\right) = e^{\hat{L}_1\Delta t}e^{\hat{L}_2\Delta t}y\left(0\right) + \mathcal{O}\left(\Delta t\right).
\end{equation}

This is equivalent to applying the separate operators in sequence. The error of this scheme is first order in the time-step. \citet{Strang1968} developed a second order scheme, of the following form:

\begin{equation} \label{eq:strang}
    y\left(\Delta t\right) = e^{\hat{L}_1\Delta t/2}e^{\hat{L}_2\Delta t}e^{\hat{L}_1\Delta t/2}y\left(0\right) + \mathcal{O}\left(\Delta t^2\right).
\end{equation}

In this way one can split two operators. In a multi-physics model, we will often have more than two terms in an operator. Consider the following differential equation:

\begin{equation}
    \frac{dy}{dt} = \sum_il_i\left(y\right).
\end{equation}

We can split this operator by partitioning terms between sub-operators $\hat{L}_1$ and $\hat{L}_2$, and recursively splitting sub-operators consisting of more than one term. This retains the second order scaling because no terms of $\mathcal{O}\left(\Delta t\right)$ are introduced. 

Partitioning between $\hat{L}_1$ and $\hat{L}_2$ can be done in a variety of ways, but the most practical is to split off one term into $\hat{L}_1$, and leave the rest in $\hat{L}_2$. This minimises the number of operator applications; if $L_1$ ever has to be split again, this has to be done twice, increasing the number of operators by four. If $L_2$ has to be split, the number of operators only increases by two. For $N$ operators, the split operator will then be of the following form:

\begin{equation}
\begin{split}
    y\left(\Delta t\right) = & e^{\hat{l}_1\Delta t/2} e^{\hat{l}_2\Delta t/2} ... e^{\hat{l}_{N-1}\Delta t/2} e^{\hat{l}_N\Delta t} \times \\
    & e^{\hat{l}_{N-1}\Delta t/2} ... e^{\hat{l}_2\Delta t/2} e^{\hat{l}_1\Delta t/2}.
\end{split}
\end{equation}

\subsection{Adaptive time operator splitting}

Strang splitting also naturally lends itself to adaptive time-stepping schemes. For this purpose, we separate operators into a {\it slow} and a {\it fast} set with respect to a given time-step $\Delta t$; a slow operator can be applied for $\Delta t$ within some required tolerance, while a fast operator can not. All slow operators can be partitioned into $\hat{L}_1$ and applied for $\Delta t$, and all fast operators can be partitioned into $\hat{L}_2$. They might not be slow operators at $\Delta t/2$, either, but we can continue splitting the operators recursively until there are no more fast operators. Successively halving the time-step results in an exponential decrease, giving the algorithm a large dynamical range in timescales.

Determining slow and fast operators is not trivial. It is not necessarily based only on how rapidly an operator works on its section of the data vector, but also on how other operators work on the same section of the data vector. A similar problem was considered by \citet{Janes2014} for Newtonian gravity n-body models. They pointed out that even if all particles evolve rapidly (they gave the example of a cluster of hard stellar binaries, which all evolve on the orbital timescale of the hard binaries), the interactions between stars in different binaries operate on timescales similar to those of single stars in the same cluster. They instead proposed an algorithm based on connected components, and showed that it reduced the wall-clock run-time of a simulation of a cluster of hard binaries by up to two orders of magnitude, compared to older adaptive time-stepping schemes.

The algorithm would decompose the particles into subsystems given a set of particles, a definition of the interaction timescale $\tau_{G,ij}$ between two particles $i$ and $j$ (e.g. the minimum of the flyby and free-fall times), and an evolution timescale $\Delta t$. Within a subsystem, or connected component, each particle $i$ would interact with at least one other particle $j$ with $\tau_{G,ij}<\Delta t$, while between two subsystems, there were no pairs $ij$ with $\tau_{G,ij}<\Delta t$. This does put particles with $\tau_{G,ij}>\Delta t$ in the same subsystem, but always because there is some path of particle pairs with timescales $<\Delta t$ connecting the two.

In the Newtonian gravity n-body model the flyby and free-fall time
were characteristic timescales that could be straightforwardly derived
from the model. For other models this may not be as simple. In {\sc
  Venice}, we leave this as a system parameter. Given $N$ operators,
this takes the form of a symmetric $N\times N$ matrix $\tau$, where
$\tau_{ij}=\tau_{ji}$ is the coupling timescale between operators
$\hat{l}_i$ and $\hat{l}_j$. Two uncoupled operators (for example,
that describe system components that don't directly interact), have a
coupling timescale of infinity; they can be applied for an arbitrarily
large time-step $\Delta t$ without having to be coupled \footnote{Note
the difference between timescales and time-steps: timescales are a
derived physical quantity of a system, while time-steps are a property
of the algorithm, external to the physics of the system.}.

Decomposition into connected components results in a set $\hat{C}_i$, where each is a set consisting of $>1$ operator (where operators within a $\hat{C}_i$ interact on timescales $<\Delta t$, but operators in different $\hat{C}_i$'s do not), and a rest set $\hat{R}$ of operators that are not connected to any other (and do not interact with operators on timescales $<\Delta t$). Each connected component $\hat{C}_i$ is fast compared to $\Delta t$, and has to be partitioned in $\hat{L}_1$. The operators in the rest set $\hat{R}$ are all slow compared to $\Delta t$, and can be partitioned into $\hat{L}_2$. This results in the following partitioning:

\begin{equation} \label{eq:cc}
    \frac{dy}{dt} = e^{\left(\sum_i\hat{C}_i\Delta t/2\right)} e^{\left(\hat{R}\Delta t\right)} e^{\left(\sum_i\hat{C}_i\Delta t/2\right)}.
\end{equation}

The operators in the rest set $R$, and the connected sets of operators $C_1..C_K$, must again be split, which can be done by linear splitting (Eq. \ref{eq:linear}) or Strang splitting (Eq. \ref{eq:strang}). Then, each individual $C_i$ can again be split into connected components. Note that if the connected components $C_i$ are split by Strang splitting, one is split at $\Delta t/2$, while the others are split at $\Delta t/4$.

In Algorithm \ref{alg:Venice_basic}, we show the basic {\sc Venice} algorithm. For simplicity we use first order splitting of the set $C_1..C_K$ and of $R$.

\begin{algorithm}
\caption{The basic {\sc Venice} algorithm, with $1^\textrm{st}$ order coupling.}
\label{alg:Venice_basic}
\begin{algorithmic}[1]
\Procedure{evolve\_cc}{$L$, $\Delta t$, $\tau$}
\LineComment{Split codes into connected components}
\State $C_1$..$C_K$, $R$ = \Call{split\_cc}{$L$, $\Delta t$, $\tau$}
\LineComment{Evolve connected components for half time}
\For{$C_i$ in $C_1$..$C_k$}
    \State \Call{evolve\_cc}{$C_i$, $\Delta t/2$, $\tau$}
\EndFor
\LineComment{Evolve rest codes for full time}
\For{$l_i$ in $R$}
    \State \Call{evolve}{$l_i$, $\Delta t$}
\EndFor
\LineComment{Evolve connected components}
\LineComment{for another half time}
\For{$C_i$ in $C_1$..$C_k$}
    \State \Call{evolve\_cc}{$C_i$, $\Delta t/2$, $\tau$}
\EndFor
\EndProcedure
\end{algorithmic}
\end{algorithm}

\begin{figure*}
    \centering
    \includegraphics[width=\linewidth]{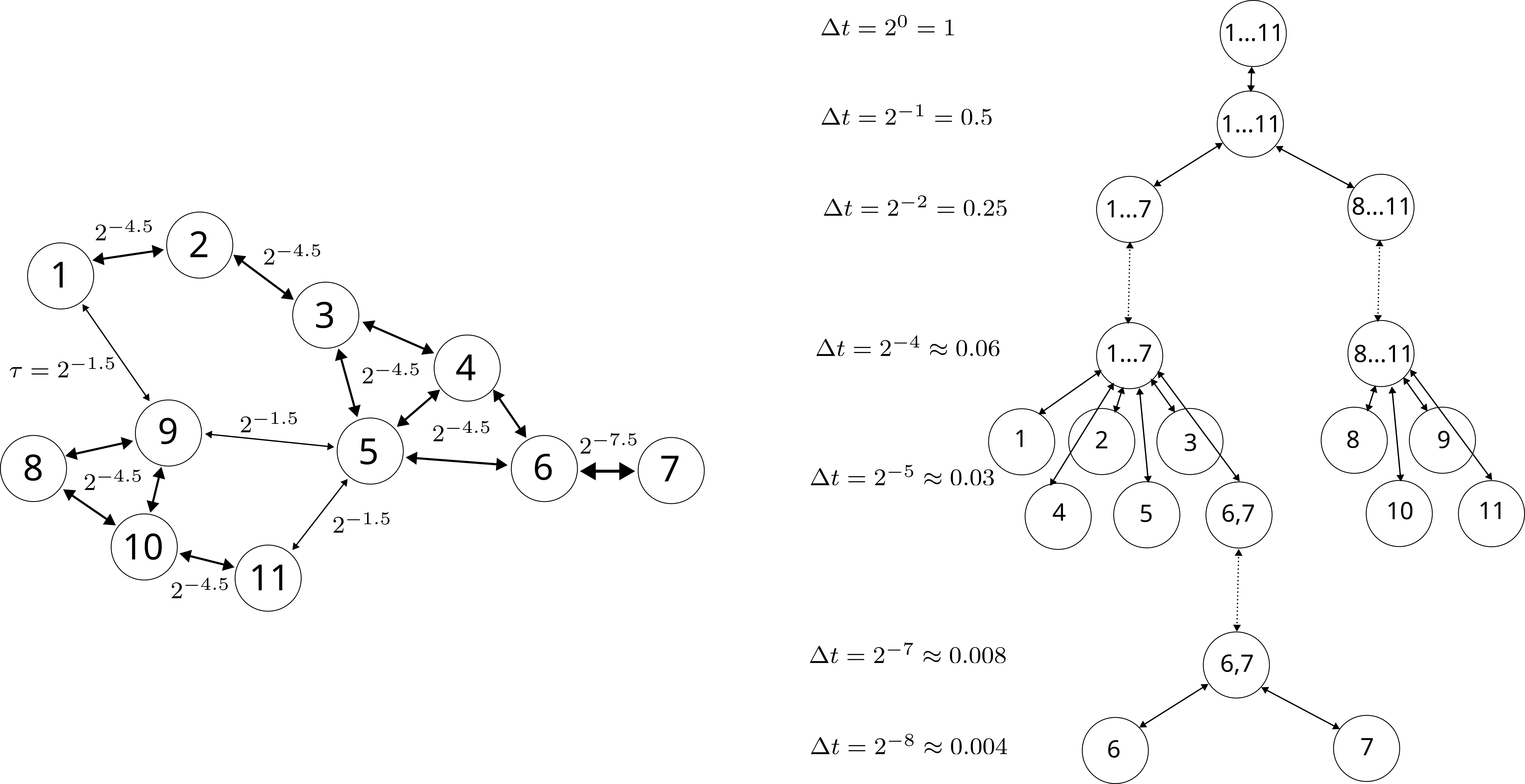}
    \caption{{\it Left:} a representation of the coupling timescale matrix $\tau$ of an operator set as an undirected, weighted graph. Edges are labelled with the coupling timescale; some edges between triangulated nodes share labels for clarity. The coupling timescale between unconnected nodes is $\infty$. {\it Right:} a tree representation of the recursive splitting of connected components, for the system on the left. Dotted lines denote skipped layers of nodes with a single parent and child node.}
    \label{fig:cctree}
\end{figure*}

In Figure \ref{fig:cctree} we show an illustration of a system of 11
operators, coupled on different (unit-less for clarity)
timescales. This system decomposes into two subsystems at $\Delta
t=2^{-2} \equiv 0.25$, meaning that the two subsystems exchange
information with an interval of $0.25$ (note that time is considered a
dimensionless quantity here). Most of the operators in these
subsystems are then split at a time-step of $\Delta t=2^{-5}\approx
0.03$, except for operators 6 and 7, which are more tightly coupled
and only split at a time-step of $\Delta t=2^{-8}\approx 0.004$.

\subsection{Kick interactions} \label{sec:kicks}

The {\sc bridge} scheme was introduced by \citet{Fujii2007} in order
to couple different gravitational systems, such as stellar clusters
within a galaxy. It uses operator splitting to split the system into
individual subsystems and interaction terms, each with their own
time-steps. In this way, a stellar cluster's internal dynamics can be
solved on a shorter timescale than the cluster's orbit in the galaxy,
or the dynamics of the galaxy. This method was expanded by
\citet{Pelupessy2012b} to also allow gravitational coupling with
smoothed-particle hydrodynamical models. \citet{PortegiesZwart2020}
expanded the method to higher-order coupling, and have introduced
augmented coupling patterns that allow for external forces (e.g. the
YORP effect and general relativity into an otherwise Newtonian
solver). We discuss how such a splitting can be incorporated into {\sc
  Venice}.

A bridged system can be seen as the following operator:

\begin{equation}
    l_i = \sum_j \left( l_i^{\left(j\right)} + \sum_k \tilde{l}_i^{\left(jk\right)} \right),
\end{equation}

\noindent where $y_i^{\left(j\right)} \cap y_i^{\left(k\right)} = 0 \quad \forall \quad j \neq k$, i.e., all $l_i^{\left(j\right)}$ use and change a different subvector of $y_i$. These are the individual n-body systems. The operators $\tilde{l}_i^{\left(jk\right)}$, however, use both $y_i^{\left(j\right)}$ and $y_i^{\left(k\right)}$, but affect only $y_i^{\left(j\right)}$. These are the kick interactions that work on n-body system $j$. 

If we split the kick interactions $\hat{\tilde{l}}_i^{\left(jk\right)}$ into $L_1$ and the evolutions $\hat{l}_i^{\left(j\right)}$ into $L_2$, we obtain the following second-order operator splitting \citep[also see][for higher-order couplings]{PortegiesZwart2020}:

\begin{equation} \label{eq:bridge}
\begin{split}
    e^{l_i\Delta t} = & \left(\prod_j\prod_k e^{\hat{\tilde{l}}_i^{\left(jk\right)}\Delta t/2}\right) \left(\prod_je^{\hat{l}_i^{\left(j\right)}\Delta t}\right) \times \\
    & \left(\prod_j\prod_k e^{\hat{\tilde{l}}_i^{\left(jk\right)}\Delta t/2}\right).
\end{split}
\end{equation}

If an operator equivalent to a bridged system is present, it can be desirable to split it explicitly for flexibility and transparency. The individual operators $l_i^{\left(j\right)}$ can be treated much in the same way as other operators $l_i$, but the kick operators $\tilde{l}_i^{\left(jk\right)}$ require special treatment. Formally they must be applied directly before and after the bridged operators, without any other, non-bridged, operator in between. However, this becomes impossible when a bridged subsystem is also coupled to another operator on a shorter timescale\footnote{Note the implicit assumption here that the {\sc bridge} timescale is the same as the previously defined coupling timescales $\tau_{ij}$}. In that case, the other operator must be included in the second factor of Equation \ref{eq:bridge}, and that factor must be decomposed in connected components following Equation \ref{eq:cc}.

The question now is where to apply kick operators. We propose to place them between the evolution of the connected components and the rest set. This preserves Strang splitting if two operators are in different connected components. This is most obvious if both are in the rest set, where two kick operators flank the evolution operators. If both are in different $C_i$'s, the kick operators are applied without evolution operators in between, which instead are applied for half a time-step before and after the kick operators. This preserves the 2$^\mathrm{nd}$ order {\sc bridge} scheme.

\subsection{Dynamic coupling timescales}

The decomposition of operators into connected components happens in
every recursive evaluation, including those that happen after some
operators have been (partly) applied. This means that, if the
timescale matrix $\tau_{ij}$ were to change as the system evolves, the
structure of the connected components can also change. In this way
{\sc Venice} can create a time-stepping scheme that adapts to the
behaviour of the system. For example, if one operator represents
stellar evolution, this operator can be coupled on long timescales so
long as all stars remain on the main sequence, on shorter timescales
when one or more stars move to more rapidly evolving stages of
evolution, and then again on longer timescales when those stars become
inert stellar remnants.

The coupled system remains consistent (with all operators applied for
the same total time) so long as the connected components within one
recursive function call of {\sc evolve\_cc} (see Algorithm
\ref{alg:Venice_basic}) are not changed (i.e. operators $l_i$ do not
move between different connected components $C_i$ and/or the rest set
$R$). However, between two subsequent recursive function calls of {\sc
  evolve\_cc} of a connected component $C_i$, the decomposition into
further connected components may be different.

In Algorithm \ref{alg:Venice_extended}, we show the {\sc Venice}
algorithm extended with kick interactions and dynamic coupling
timescales. For clarity, we present the algorithm with first-order
splitting of $R$ and the set $C_1..C_K$.

\begin{algorithm*}
\caption{The extended {\sc Venice} algorithm, with $1^\textrm{st}$ order coupling.}
\label{alg:Venice_extended}
\begin{algorithmic}[1]
\Procedure{evolve\_cc}{$L$, $\Delta t$, $\tau$}
\LineComment{Update coupling timescales}
\State \Call{update\_timescales}{L, $\Delta t$, $\tau$}
\LineComment{Split codes into connected components}
\State $C_1$..$C_K$, $R$ = \Call{split\_cc}{$L$, $\Delta t$, $\tau$}
\LineComment{Evolve connected components for half time}
\For{$C_i$ in $C_1$..$C_k$}
    \State \Call{evolve\_cc}{$C_i$, $\Delta t/2$, $\tau$}
\EndFor
\LineComment{Kick between connected components for half time}
\For{$l_i$ in $l$}
    \For{$l_j$ in $l$}
        \If{$l_i$ and $l_j$ not in same $C_k$}
            \State \Call{kick}{$l_i$, $l_j$, $\Delta t/2$}
        \EndIf
    \EndFor
\EndFor
\LineComment{Evolve rest codes for full time}
\For{$l_i$ in $R$}
    \State \Call{evolve}{$l_i$, $\Delta t$}
\EndFor
\LineComment{Kick between connected components for another half time}
\For{$l_i$ in $l$}
    \For{$l_j$ in $l$}
        \If{$l_i$ and $l_j$ not in same $C_k$}
            \State \Call{kick}{$l_i$, $l_j$, $\Delta t/2$}
        \EndIf
    \EndFor
\EndFor
\LineComment{Evolve connected components for another half time}
\For{$C_i$ in $C_1$..$C_k$}
    \State \Call{evolve\_cc}{$C_i$, $\Delta t/2$, $\tau$}
\EndFor
\EndProcedure
\end{algorithmic}
\end{algorithm*}

\subsection{{\sc amuse} implementation}

Thus far we have discussed {\sc Venice} as an operator splitting algorithm. However, it can also be used as a coupling scheme for physical models that may not strictly solve a proper partial differential equation. The {\sc amuse} platform provides a standardised structure for (interfaces to) physical models, making it an excellent environment to develop a reference implementation of the algorithm.

In the previous sections, we tacitly assumed that all operators affect the same data vector. However, in {\sc amuse} every code has its own data vector (typically in the form of particle or grid data). Coupling must be done explicitly, by copying data from one code to another. In this implementation we use the channels in {\sc amuse}. These are defined from one data structure to another, and can copy any subset of data from one structure to the other. They must be explicitly called, which we do between operations. We could have chosen a lazy approach, where before the evolution of a code we copy all channels to that code. However, this results in an end state where the data of different codes is different, and it is non-trivial to reconstruct which code contains the most recent state. Instead, we take an active approach, where we copy all channels from a code after its evolution step. This can introduce unnecessary copy operations, but is less prone to mistakes. 

In many situations the structure of a code's data set will be updated throughout the simulation (e.g. adaptive mesh refinement codes, particle-based stellar winds and sink particles). To resolve this, every pair of codes can have a function that synchronises the data sets, defined by the user. This is done before channels are copied to ensure the data structures are compatible.

We also implemented a data output framework. This allows for analysis of fast subsystems without needing the same short time-step for the entire system (or the construction of a wrapper class whose only purpose besides evolving is writing data). We write three types of output: checkpoint files, plot files, and debug files. Checkpoint files are saved after the evolution of all codes, and contain the state of the entire system at a synchronised time. Plot files are saved after the evolution of an entire connected component. To prevent redundant files, saving checkpoint and plot files is skipped if there is only a single connected component, consisting of more than one code; in that case, the state at the end of the function is the same as at the end of the second recursive call. Debug files are saved after the evolution of individual codes. 

The {\sc amuse} implementation is available on Github: \url{https://github.com/amusecode/venice/}.

\section{Testing {\sc Venice}}

In this section we present a number of tests of {\sc Venice} using the {\sc amuse} implementation. The scripts to run these tests, and produce the figures, are also available on the {\sc Venice} Github page.

\subsection{Gravity and stellar evolution}

Coupled gravity-stellar evolution simulations are one of the oldest types of astrophysical multi-physics simulations, going back to \citet{Terlevich1987}. The coupling of these systems is relatively straightforward. Assuming a collisionless system and no changes in momentum during stellar evolution (e.g. supernova kicks), the only coupling is through the mass. Stars lose mass in stellar winds and at the end of their lifetime, and this changes the gravitational acceleration they induce in other stars. Newton's equations of motion can account for changing mass in differential form, but most astrophysical gravitational dynamics models do not include this. Coupling is typically done through operator splitting.

We use the {\sc ph4} \citep{McMillan2012} gravitational dynamics code and the {\sc SeBa} parameterised stellar evolution code \citep{PortegiesZwart1996,Toonen2012}. Coupling is done through a channel copying mass from the stellar evolution code to the gravitational dynamics code.

In Section \ref{sec:grav_stel_const} we use a constant coupling timescale. This becomes problematic if the stellar evolution timescale changes. This happens, for example, when a star leaves the main sequence. In Section \ref{sec:grav_stel_adapt}, we expand the model with a dynamic coupling timescale.

For the initial conditions we adopt a 100 star Plummer sphere in virial equilibrium. Stellar masses are distributed following a Kroupa IMF \citep{Kroupa2001}, between 0.08 and 8 M$_\odot$. We manually set a randomly selected star to a higher mass, 16 M$_\odot$ in Section \ref{sec:grav_stel_const} and 50 M$_\odot$ in Section \ref{sec:grav_stel_adapt}. The Plummer radius is set such that the free-fall time is 6 Myr (resulting in a Plummer radius of $\sim$1.25 pc with the 16 M$_\odot$ star, and $\sim$1.47 pc with the 50 M$_\odot$ star). This is also the total run-time of the tests. During this time, the 16 M$_\odot$ star remains on the main sequence, and loses relatively little mass, allowing us to use relatively long coupling timescales compared to the total simulation time. The 50 M$_\odot$ star leaves the main sequence and undergoes a supernova explosion during this time frame. This results in more rapid changes in mass, requiring very short coupling timescales at certain epochs. In Section \ref{sec:grav_stel_adapt}, we demonstrate a timescale update function that can shorten the coupling timescale appropriately.

\subsubsection{Constant coupling timescale} \label{sec:grav_stel_const}

In Algorithm \ref{alg:gravity_stellar_constant} we show how a {\sc Venice} system of gravitational dynamics and stellar evolution coupled at a constant timescale is set up.

\begin{algorithm}
\caption{{\sc Venice} setup of gravitational dynamics and stellar evolution coupled at a constant timescale}
\label{alg:gravity_stellar_constant}
\begin{algorithmic}[1]
\Procedure{setup\_gravity\_stellar\_constant}{ gravity, stellar, dt}
\State system = \Call{Venice}{\null}
\State system.\Call{add\_code}{gravity}
\State system.\Call{add\_code}{stellar}
\State system.timescale\_matrix[0][1] = dt
\LineComment{Copy mass from code 1 (stellar)}
\LineComment{to 0 (gravity)}
\State system.\Call{add\_channel}{1, 0, ["mass"]}
\State return system
\EndProcedure
\end{algorithmic}
\end{algorithm}

\begin{figure*}
    \centering
    \includegraphics[width=\linewidth]{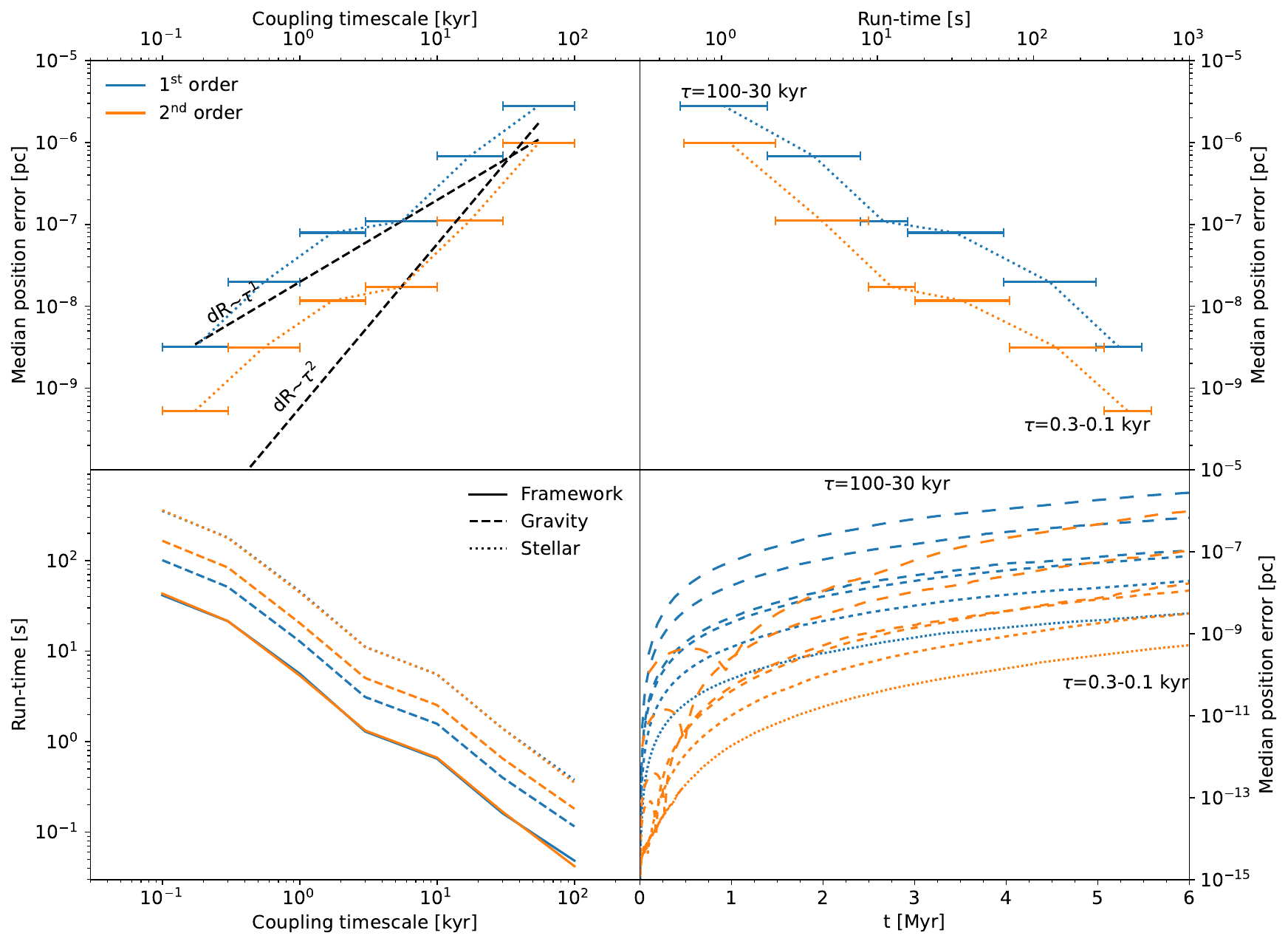}
    \caption{Convergence and performance of a gravitational dynamics and a stellar evolution model coupled at a constant timescale. The top panels and bottom left panel show three views of the median position error, coupling timescale, and wall-clock run-time as functions of each other. The bottom right panel shows the time evolution of the median position error in all runs. The black dashed lines in the top left plot show linear and quadratic scalings of the error with the timescale.}
    \label{fig:gravity_stellar_constant_scaling}
\end{figure*}

In Figure \ref{fig:gravity_stellar_constant_scaling} we show three views of the median position error, coupling timescale, and wall-clock run-time as functions of each other, for both 1$^\mathrm{st}$ and 2$^\mathrm{nd}$ order splitting of the operator rest set $R$ (the splitting order of the connected components $C_1..C_K$ does not affect the splitting procedure for $<4$ operators). We also show the time evolution of the median position error. The position error is defined as the distance between initially identical particles in runs with different timescales. We compute position errors between subsequent values of the coupling timescale, indicated by the end points of error bars. 

Most gravitational systems of more than two bodies are chaotic \citep{Miller1964}. As a result, any perturbation we introduce in the system triggers an exponential growth of that perturbation. Changing the timescale from one run to another is such a perturbation, and therefore triggers exponential growth. We observe that the median position error grows roughly linearly with time (although this may be the first-order behaviour of exponential growth at times much smaller than the Lyapunov time), and that there is a systematic offset between curves associated with subsequent coupling timescales. Therefore, we argue that the systematic change is caused by the coupling timescale, and not by the chaotic behaviour of the system.

The median position error scales approximately linearly with the coupling timescale, for both 1$^\mathrm{st}$ and 2$^\mathrm{nd}$ order coupling. However, at the same time-step the error of the $2^\textrm{nd}$ order method is an order of magnitude smaller. This comes at the cost of slightly longer wall-clock run-time\footnote{The increase in run-time is due to the gravity model running twice as long.} but the median position error at a given run-time is typically lower for $2^\textrm{nd}$ order than for $1^\textrm{st}$ order. In this model configuration (where the gravity model is the first element of the operator list $L$), it is the gravity model that is evolved in two steps due to Strang splitting. This can be explained by the gravity model's preferred internal time-step being larger than the largest coupling timescale, so that it performs a single numerical step every time {\sc Venice} evolves the model. In this case, the gravity model does not dominate computation time, so the increase in run-time is small. This implies that $2^\textrm{nd}$ order coupling results in the best trade-off between run-time and performance unless both models have comparable run-time and have to shorten their internal time-step.

\begin{figure}
    \centering
    \includegraphics[width=1\linewidth]{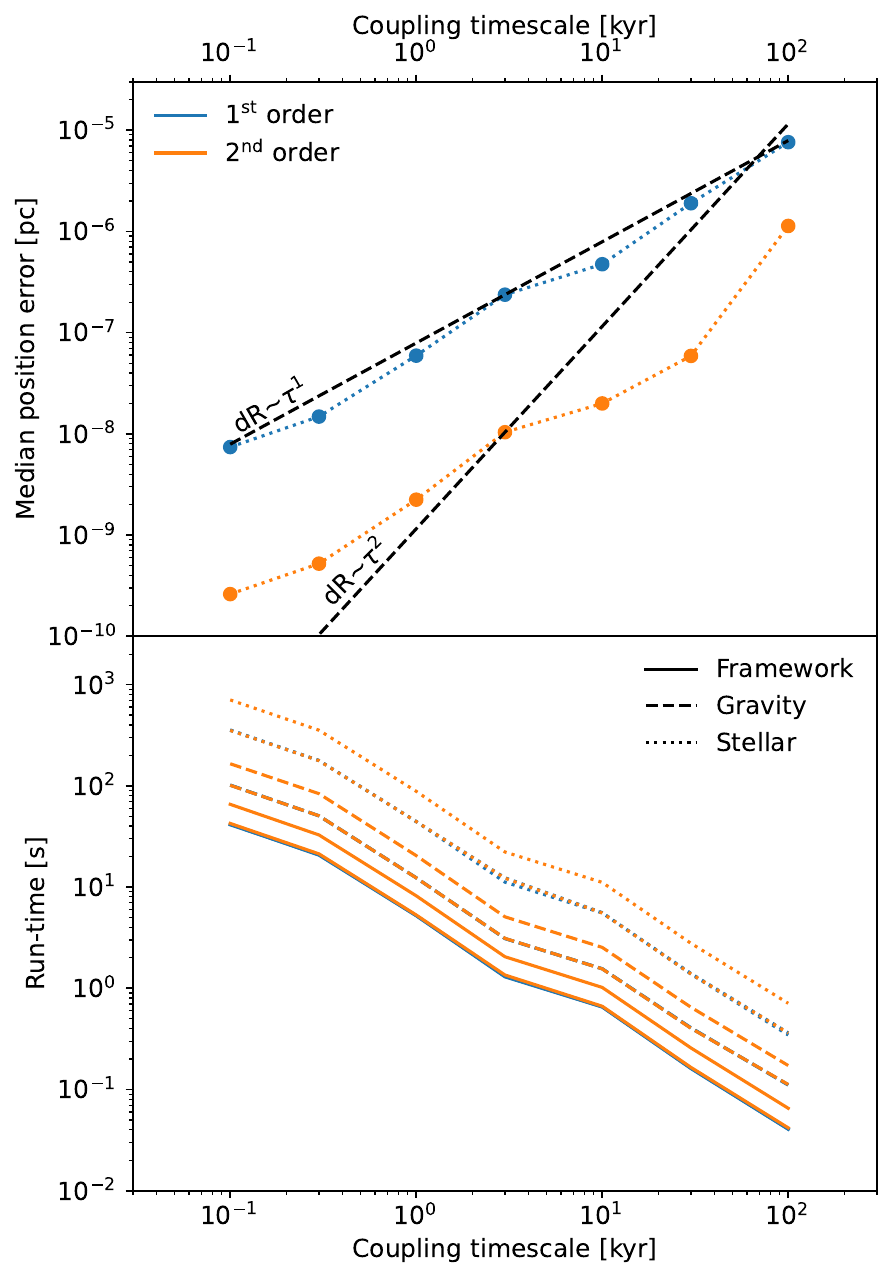}
    \caption{Convergence and performance of a gravitational dynamics and a stellar evolution model coupled at a constant timescale, with reversed operators. We show, as a function of coupling timescale, the median position error (top) and wall-clock run-time (bottom). The black dashed lines in the top plot show linear and quadratic scalings of the error with the timescale.}
    \label{fig:gravity_stellar_reversed_scaling}
\end{figure}

The order in which operators are split affects the outcome of numerical integration. To quantify this effect we compare the model above with a model that is identical, except that the stellar evolution model is the first operator in $L$. This is done by reversing the order of {\sc add\_code} calls in Algorithm \ref{alg:gravity_stellar_constant}, and results in the stellar evolution model being evolved twice per Strang splitting, instead of the gravity model. In Figure \ref{fig:gravity_stellar_reversed_scaling}, we show how the median position error and the wall-clock run-time depend on the coupling timescale, for the two reversed models, at the same coupling timescale, for both 1$^\mathrm{st}$ and 2$^\mathrm{nd}$ order coupling. 

We observe globally the same behaviour as for the convergence with timescale; the median position errors converges linearly with the timescale, for both 1$^\mathrm{st}$ and 2$^\mathrm{nd}$ order coupling, although 2$^\mathrm{nd}$ order coupling results in smaller errors. The magnitude of the errors are also roughly the same at a given coupling timescale. Finally, we again see that the code being evolved twice has a longer run-time\footnote{If the system is evolved for multiple time-steps $\Delta t$, there may be two sequential half-time-steps, e.g.: $e^{\hat{L}_1\Delta t/2}e^{\hat{L}_2\Delta t}\left(e^{\hat{L}_1\Delta t/2}e^{\hat{L}_1\Delta t/2}\right)e^{\hat{L}_2\Delta t}e^{\hat{L}_1\Delta t/2}$. These could be combined in a single evolution step, with a potential for faster evolution, but for larger systems there could be operators in between, and the coupling timescales may change in between. Therefore such recombination is not possible.}. This shows that the order of operators impacts the efficiency of the model, and while it also impacts the results, this error converges with shorter coupling timescales.

\subsubsection{Adaptive coupling timescale} \label{sec:grav_stel_adapt}

In the test above, we were able to achieve a numerically converging solution at relatively long coupling timescales because no star evolved off the main sequence during the integration time. Off the main sequence, stars evolve rapidly, ultimately going supernova on a stellar free-fall time, or shedding their envelopes as planetary nebulae. Evolving the entire system at the shortest possible timescale at which the stellar population can evolve would result in a large overhead, leading to many unnecessary couplings. {\sc Venice} can easily resolve this using an {\sc update\_timescale} function. Setting this to some fraction $\eta$ of the evolution timescale of the fastest evolving star allows the system to couple more frequently when needed.

In principle, the most efficient integration would be achieved by coupling every star on its own timescale; the scheme we described evolves all stars, including those on the main sequence, on the timescale of the star with the shortest time-step. However, this increases the overhead of {\sc Venice} (note that the connected component algorithm scales quadratically with the number of operators). The optimal solution is to separate stars into a number of mass bins, but calibrating the optimal method of distribution is beyond the scope of this work.

In Algorithm \ref{alg:gravity_stellar_adaptive} we show how a {\sc Venice} system of gravitational dynamics and stellar evolution coupled at a dynamic timescale is set up.

\begin{algorithm}
\caption{{\sc Venice} setup of gravitational dynamics and stellar evolution coupled at an adaptive timescale}
\label{alg:gravity_stellar_adaptive}
\begin{algorithmic}[1]
\Procedure{setup\_gravity\_stellar\_adaptive}{ gravity, stellar, $\eta$}
\State system = \Call{Venice}{\null}
\State system.\Call{add\_code}{gravity}
\State system.\Call{add\_code}{stellar}
\LineComment{\textbf{lambda} defines a function, as in Python}
\State system.update\_timescale[0][1] = 
\State \hskip1.5em \textbf{lambda} code1, code2, dt: 
\State \hskip1.5em $\eta$ * code2.stellar\_timesteps.\Call{min}{\null}
\LineComment{Copy mass from code 1 (stellar)}
\LineComment{to 0 (gravity)}
\State system.\Call{add\_channel}{1, 0, ["mass"]}
\State return system
\EndProcedure
\end{algorithmic}
\end{algorithm}

We use the same initial conditions as in the previous section, except with a 50 M$_\odot$ star instead of a 16 M$_\odot$ star, and rescaling the cluster radius and stellar velocities such that the dynamical time is still 6 Myr.

\begin{figure}
    \centering
    \includegraphics[width=1\linewidth]{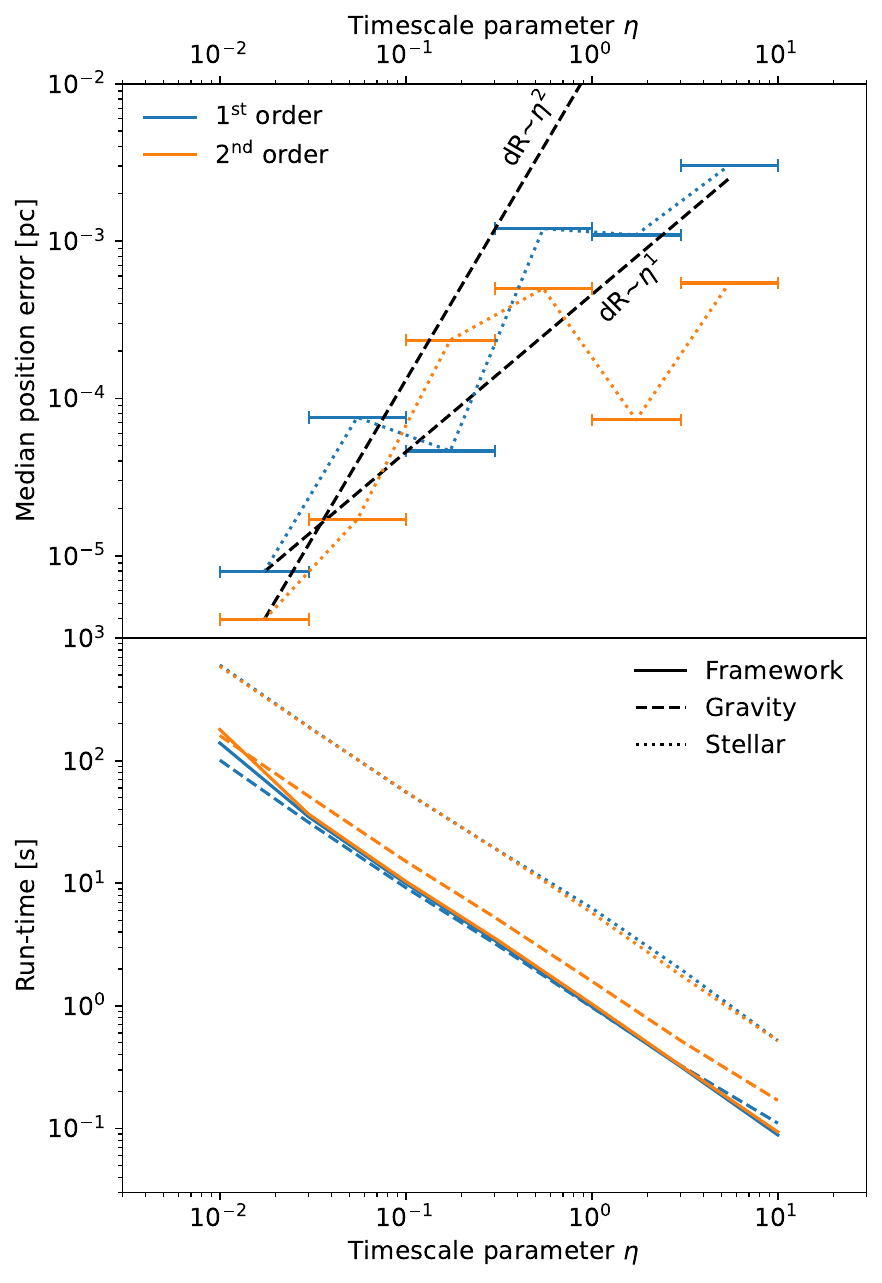}
    \caption{Convergence and performance of a gravitational dynamics and a stellar evolution model coupled at an adaptive timescale. We show, as a function of the timescale parameter $\eta$, the median position error (top) and wall-clock run-time (bottom). The median position error is computed for two subsequent values of the coupling timescale, indicated by the endpoints of the error bars. The black dashed lines in the top plot show linear and quadratic scalings of the error with the timescale.}
    \label{fig:gravity_stellar_adaptive_scaling}
\end{figure}

In Figure \ref{fig:gravity_stellar_adaptive_scaling} we show the median position error (top) and the wall-clock run-time (bottom) as a function of the coupling timescale parameter $\eta$, for both 1$^\mathrm{st}$ and 2$^\mathrm{nd}$ order coupling. The median position error scales roughly linearly for both 1$^\mathrm{st}$ and 2$^\mathrm{nd}$ order coupling, but the error does not monotonically scale with the timescale parameter. The increase in run-time at $2^\mathrm{nd}$ order coupling is again due to gravity being split into two evolution steps.

\begin{figure*}
    \centering
    \includegraphics[width=1\linewidth]{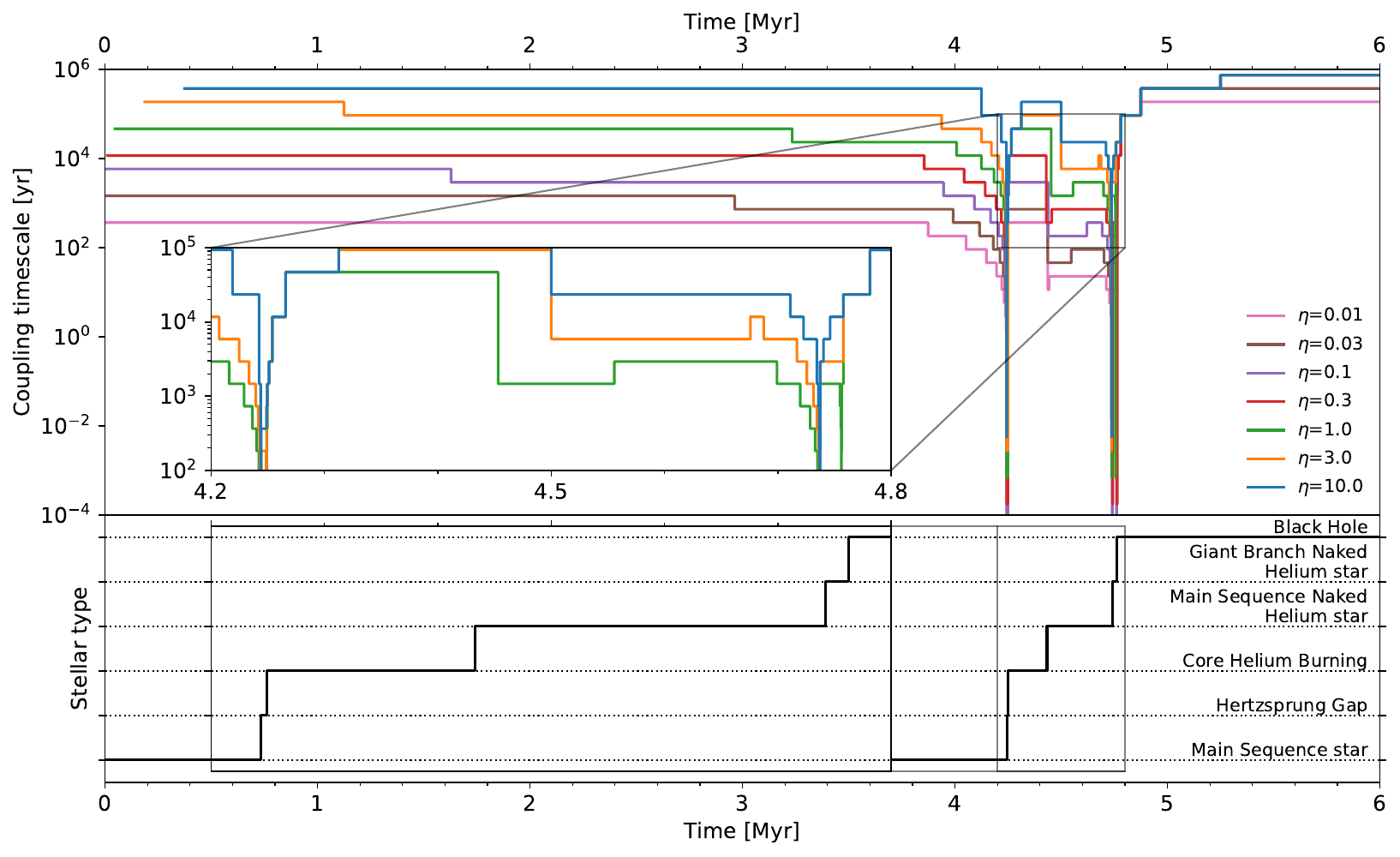}
    \caption{{\it Top:} The dynamic coupling timescale between the gravitational dynamics and stellar evolution models as a function of time, for a number of timescale parameters $\eta$. {\it Bottom:} The stellar type through time of a 50 M$_\odot$ star, which is the most massive one in the system of the top panel.}
    \label{fig:coupling_timescale_evolution}
\end{figure*}

In Figure \ref{fig:coupling_timescale_evolution} we show how the
coupling timescale evolves as a function of time for a number of
values of $\eta$ (top), and the stellar type of the 50 M$_\odot$ star
(bottom) over the same time interval. Over the first 4 Myr, the star
is on the main sequence and the coupling timescale is mostly
constant. At a number of transitions between phases (before and after
the Hertzsprung gap and giant branch naked helium star phases,
sometimes called ``stripped stars'', these days), the coupling
timescale drops by up to seven orders of magnitude, but also quickly
increase by orders of magnitude afterwards. Then, when the star is a
black hole, the coupling timescale becomes the same for values of
$\eta \ge 0.1$. This indicates that the coupling timescale dictated by
the stellar evolution model is large compared to the remaining
integration time.

We note that for $\eta>1$, there is a tendency for the model to `overshoot' phases of stellar evolution. Ideally, the ratio of coupling timescales for two values of $\eta$ should always be about $\eta$ itself, but during e.g. the transition from the Hertzsprung Gap to Core Helium burning for $\eta=10$, and the transition from Giant Branch Naked Helium Star to Black Hole for $\eta=3,10$, the sharp drop to coupling timescales $\ll$1 kyr does not occur, while for higher values of $\eta$ it does. In these cases, the stellar evolution model evolves through the rapid evolution phase without recomputing the coupling timescale. This implies that values of $\eta\lesssim 1$ should be used.

\subsection{{\sc Bridge}d gravity}

We demonstrate the {\sc bridge} algorithm in {\sc Venice} by simulating the dynamics of a star cluster in a galaxy. 

The cluster consists of 100 stars, with masses drawn from a Kroupa IMF between 0.08 and 8 M$_\odot$, distributed following a Plummer distribution with a radius of 10 pc. We again use {\sc ph4} to model the cluster's gravitational dynamics.

We couple this cluster to the analytic \citet{Bovy2015} Milky-Way-like galaxy potential, which consists of a bulge (power-law, exponentially cut-off density profile), disk \citep[Miyamoto-Nagai density profile,][]{Miyamoto1975}, and halo \citep[NFW density profile,][]{Navarro1996}.

We initialise the cluster at 8 kpc from the centre of the galaxy, with
a velocity perpendicular to the direction to the centre. We
parameterise this velocity in terms of an effective eccentricity: the
velocity is that of a Keplerian orbit, at apocenter, for a central
mass equal to the total mass within a circular orbit, given the
effective eccentricity. If this eccentricity is 0, the orbit will be
circular, but larger effective eccentricities do not result in
elliptic orbits (because the mass enclosed in the orbit is not
constant). We evolve the system for $223$\,Myr, one period of the
adopted orbit.

\subsubsection{Classical {\sc bridge}}

In Algorithm \ref{alg:bridge_constant} we show how a {\sc Venice} system of two bridged gravitational dynamics models, coupled at a constant timescale, is set up. In this case, code 0 is the galaxy potential, and code 1 is the cluster gravity.

We set the cluster's orbital velocity such that the effective eccentricity is 0. In these tests, we use the cluster's centre of mass to investigate the system's convergence.

\begin{algorithm}
\caption{{\sc Venice} setup of two gravitational systems, where code 0 kicks code 1 through {\sc bridge}, at a constant timescale.}
\label{alg:bridge_constant}
\begin{algorithmic}[1]
\Procedure{setup\_bridged\_gravity\_constant}{ gravity1, gravity2, dt}
\State system = \Call{Venice}{\null}
\State system.\Call{add\_code}{gravity1}
\State system.\Call{add\_code}{gravity2}
\State system.timescale\_matrix[0][1] = dt
\LineComment{Kick interactions from code 0 to code 1}
\State system.kick[0][1] = dynamic\_kick
\State return system
\EndProcedure
\end{algorithmic}
\end{algorithm}

\begin{figure}
    \centering
    \includegraphics[width=1\linewidth]{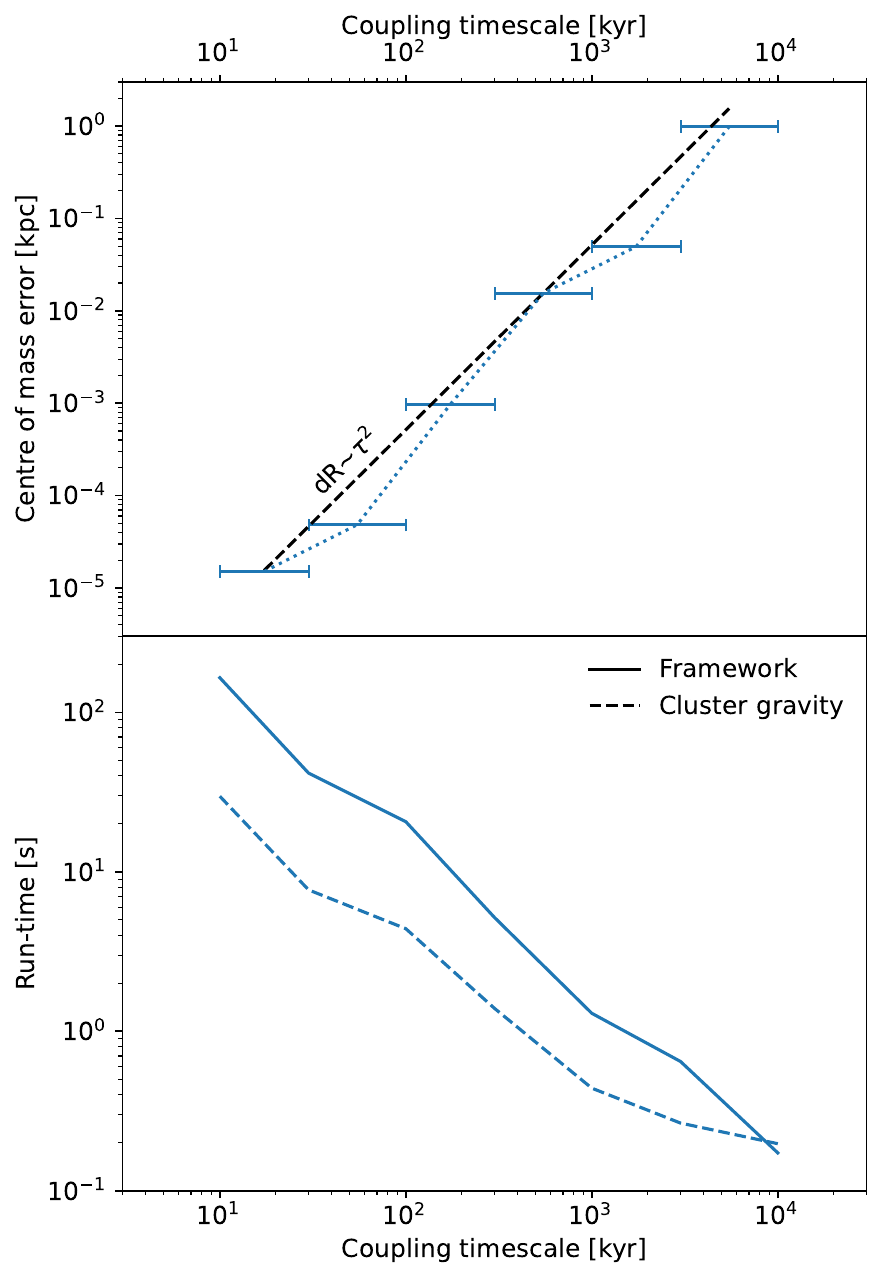}
    \caption{Centre of mass error (top) and wall-clock run-time (bottom) as a function of constant coupling timescale, for a star cluster in a background galaxy potential. Centre of mass errors are computed between subsequent values of the coupling timescale, indicated by the ends of the error bars. The dashed black line indicates quadratic scaling of the centre of mass error with the coupling timescale.}
    \label{fig:bridge_constant}
\end{figure}

In Figure \ref{fig:bridge_constant} we show how the centre of mass error (top) and wall-clock run-time (bottom) depend on the coupling timescale. Because the only coupling is done through kick interactions (and there is no shared data between the gravity codes), the splitting order of the operator rest set $R$ should not affect the result. For this reason, we only show the result of $1^\textrm{st}$ order splitting. However, we note that by construction the {\sc bridge} scheme has $2^\textrm{nd}$ order error behaviour.

The position error indeed scales quadratically with the coupling timescale. Towards long coupling timescales, the run-time of the cluster gravity model appears to move towards an asymptote. This is the point at which the code no longer requires shortened time-steps to fit within the coupling timescale, but can take large internal time-steps.

\subsubsection{Adaptive {\sc bridge}} \label{sec:bridge_adaptive}

Analogous to the time-step in the general n-body problem, the bridge time-step can potentially cover a large range. If a star cluster's orbit within a galaxy is not circular, the appropriate bridge time-step can be smaller close to the galactic centre than farther away. 

\begin{algorithm}
\caption{{\sc Venice} setup of two gravitational systems, where system 1 kicks system 2 through {\sc bridge}, at an adaptive timescale.}
\label{alg:bridge_adaptive}
\begin{algorithmic}[1]
\Procedure{setup\_bridged\_gravity\_adaptive}{ gravity1, gravity2, $\eta$}
\State system = \Call{Venice}{\null}
\State system.\Call{add\_code}{gravity1}
\State system.\Call{add\_code}{gravity2}
\State system.update\_timescale[0][1] =
\State \hskip1.5em \textbf{lambda} code1, code2, dt: 
\State \hskip1.5em $\eta$ * \Call{min\_freefall\_time}{code1, code2}
\LineComment{Kick interactions from code 0 to code 1}
\State system.kick[0][1] = dynamic\_kick
\State return system
\EndProcedure
\end{algorithmic}
\end{algorithm}

We base the adaptive coupling timescale on the freefall times of the particles in one model in the potential of the other model. The freefall time between two particles $i$ and $j$ is $\sqrt{r_{ij}/a_{ij}}$, where $r_{ij}$ is the distance between the particles, and $a_{ij}$ is the gravitational acceleration. The acceleration is computed for the {\sc bridge} interaction, and the distance is that to the centre of the central potential. Similar to the dynamic stellar coupling timescale, we now take the minimum among all particles and apply a scaling factor $\eta$. We use the initial conditions described above, and set the cluster's velocity such that the effective eccentricity is 0.5.

\begin{figure}
    \centering
    \includegraphics[width=1\linewidth]{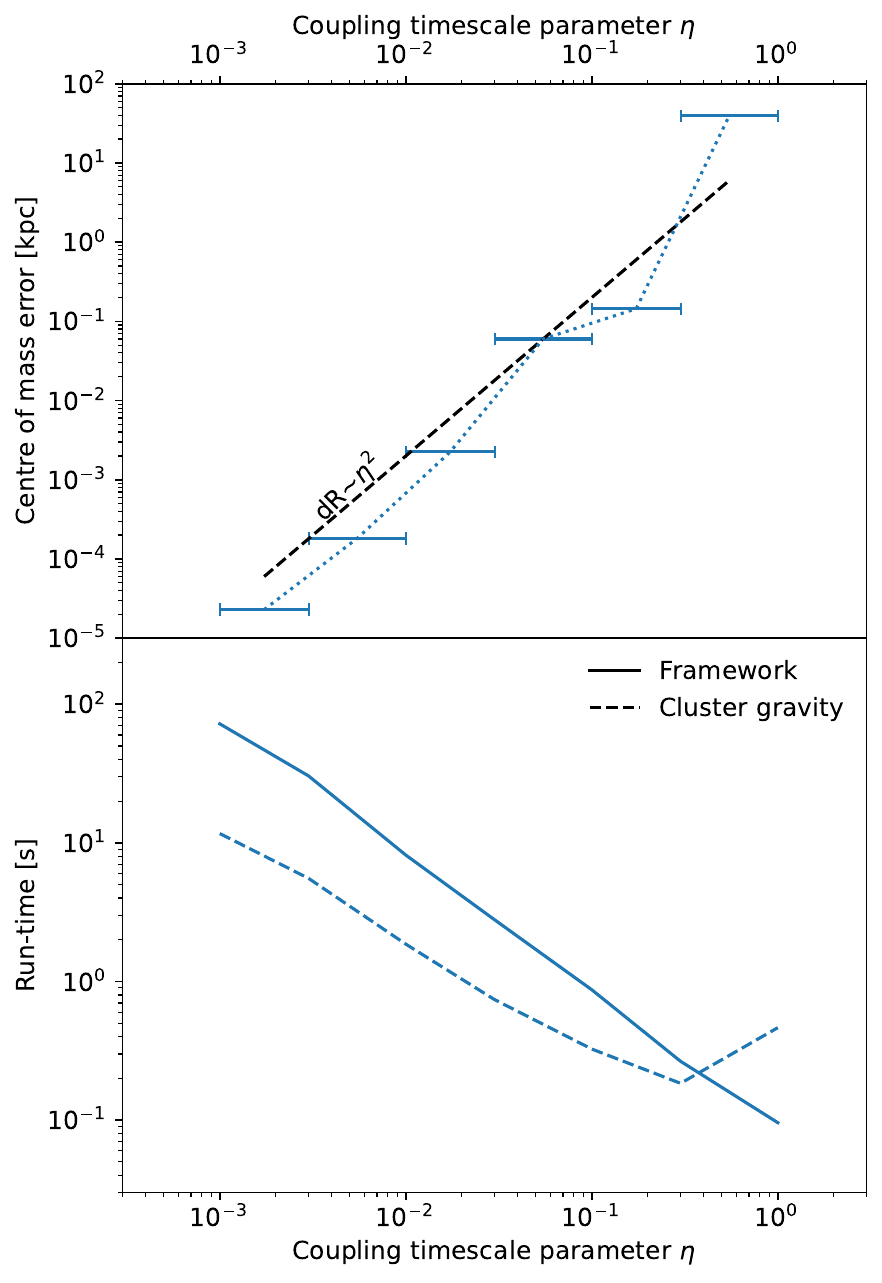}
    \caption{Centre of mass error (top) and wall-clock run-time (bottom) as a function of timescale parameter $\eta$, for a star cluster in a background galaxy potential. Centre of mass errors are computed between subsequent values of the coupling timescale, indicated by the ends of the error bars. The dashed black line indicates quadratic scaling of the centre of mass error with the coupling timescale.}
    \label{fig:bridge_adaptive}
\end{figure}

In Figure \ref{fig:bridge_adaptive} we show how the centre of mass error (top) and wall-clock run-time (bottom) depend on the coupling timescale parameter $\eta$. Similarly to the case of the constant coupling timescale, the centre of mass error converges quadratically with the coupling timescale parameter, and the cluster gravity run-time no longer decreases at large coupling timescales. We note that the centre of mass error between $\eta=0.3$ and $\eta=1$ is larger than the initial distance from the galactic centre, while between $\eta=0.1$ and $\eta=0.3$ it is almost two orders of magnitude smaller than the initial distance. This implies that a value $\eta<1$ should be used. 

\subsection{Combined model}

Having demonstrated {\sc Venice}'s capabilities in small, controlled experiments, we now demonstrate its ability to build up complex simulations. In this section we couple the models described above. We aim to simulate the dynamics of a star cluster in the potential of a galaxy, while following stellar mass loss due to stellar evolution. We set up the interactions between the cluster's gravitational dynamics and stellar evolution models adaptively, in the same way as in Section \ref{sec:grav_stel_adapt}, and the interaction between the cluster's gravitational dynamics and the background potential as in Section \ref{sec:bridge_adaptive}. We use a stellar coupling timescale parameter ($\eta_s$) of 0.3, and a gravitational coupling timescale parameter ($\eta_g$) of 0.03. 

In addition to combining the two models, we add two massive particles just outside the galactic disk, comparable in mass and location to the Large and Small Magellanic Clouds. We add these in their own gravity code, which is also coupled to the galaxy, and also perturbs the cluster (the cluster has no effect on other components). This separation into two codes allows us to couple the galaxy and these perturbers to the cluster on different timescales. In Algorithm \ref{alg:kitchen_sink}, we show how this systems is set up. 

For the cluster, we use the same 100 star, 10 pc Plummer sphere, with masses from a Kroupa IMF up to 8 M$_\odot$, except that we give three stars masses of 16, 30, and 50 M$_\odot$. The cluster is initially on a circular orbit at 5 kpc from the galactic centre. The two perturbers have a mass of $10^{11}$ M$_\odot$ each, and are initially at 50 and 60 kpc circular orbits, 180$^\circ$ and 90$^\circ$ ahead of the cluster, respectively. We evolve the system for 5 Gyr.

\begin{algorithm*}
\caption{{\sc Venice} setup of a cluster with stellar evolution in a galaxy perturbed by dwarf galaxies.}
\label{alg:kitchen_sink}
\begin{algorithmic}[1]
\Procedure{setup\_bridged\_gravity}{gravity\_cluster, stellar, gravity\_galaxy, gravity\_perturbers, $\eta_s$, $\eta_g$}
\State system = \Call{Venice}{\null}
\State system.\Call{add\_code}{gravity\_cluster}
\State system.\Call{add\_code}{stellar}
\State system.\Call{add\_code}{gravity\_galaxy}
\State system.\Call{add\_code}{gravity\_perturbers}
\State system.\Call{add\_channel}{1, 0, ["mass"]}
\LineComment{All gravity systems are coupled; stellar evolution only to cluster gravity}
\State system.update\_timescale[0][1] = \textbf{lambda} code1, code2, dt: $\eta_s$*code2.stellar\_timesteps.\Call{min}{\null}
\State system.update\_timescale[0][2] = \textbf{lambda} code1, code2, dt: $\eta_g$*\Call{min\_freefall\_time}{code1, code2}
\State system.update\_timescale[0][3] = \textbf{lambda} code1, code2, dt: $\eta_g$*\Call{min\_freefall\_time}{code1, code2}
\State system.update\_timescale[2][3] = \textbf{lambda} code1, code2, dt: $\eta_g$*\Call{min\_freefall\_time}{code1, code2}
\LineComment{Galaxy and perturbers both kick cluster}
\State system.kick[2][0] = dynamic\_kick
\State system.kick[3][0] = dynamic\_kick
\LineComment{Galaxy kicks perturbers}
\State system.kick[2][3] = dynamic\_kick
\State return system
\EndProcedure
\end{algorithmic}
\end{algorithm*}

\begin{figure*}
    \centering
    \includegraphics[width=1\linewidth]{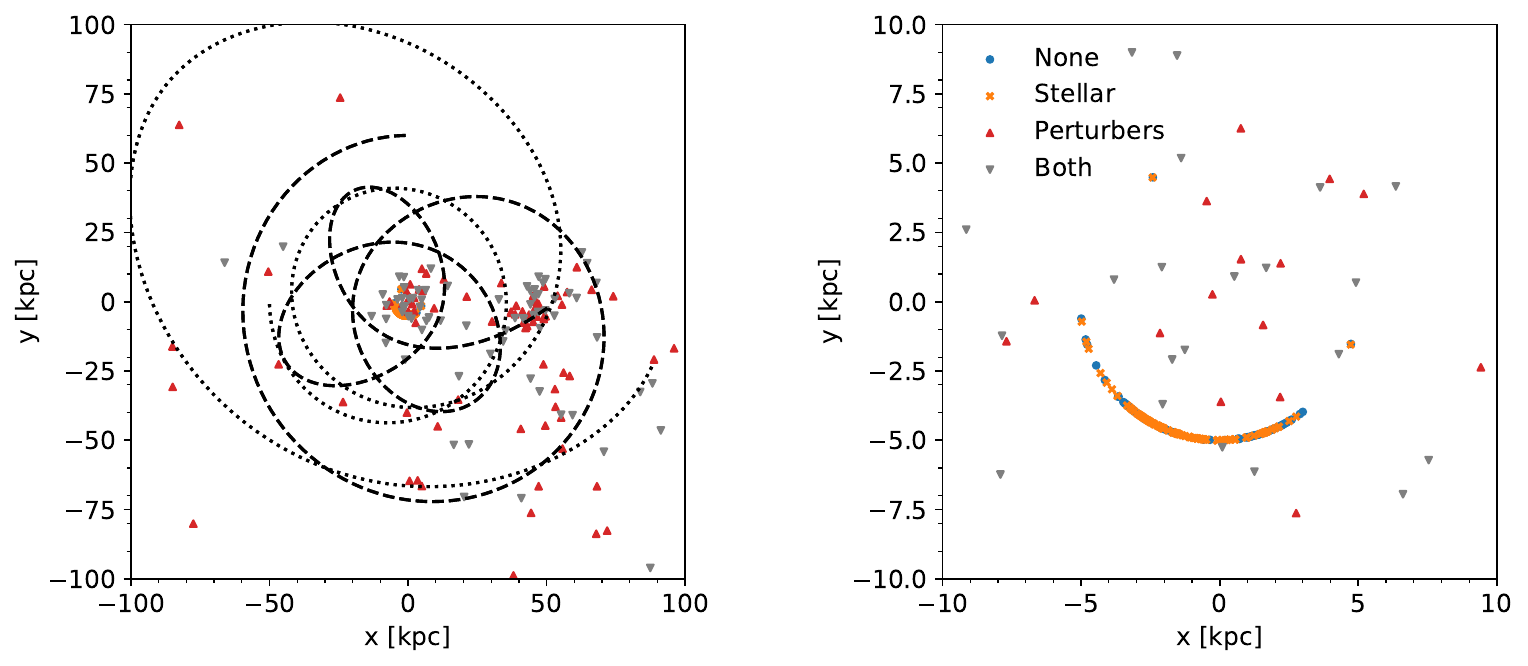}
    \caption{Maps of the combined simulation projected on the galactic plane. The left panel shows a wide view, while the right panel is zoomed in on the centre. The black dashed and dotted lines show the trajectories of the two perturbers. Points show the positions at 5 Gyr of the cluster stars, where the colour indicates the model setup (red and grey are with perturbers, orange and grey are with stellar evolution).}
    \label{fig:combined_map}
\end{figure*}

In Figure \ref{fig:combined_map}, we show projections of this system on the galactic plane. The perturbers interact before having completed a full orbit, kicking the outer body inwards (towards the inner 10-15 kpc, still outside the cluster orbit), and kicking the inner body outwards. However, the cluster is still clearly perturbed; in runs without perturbers, the cluster disperses along a segment of a circle still 5 kpc in radius, while in runs with perturbers, stars are scattered throughout the galaxy. Overdensities remain in the inner 5 kpc, and around the final position of the perturber that was scattered inwards. A number of former cluster members appear to have become bound to this perturber.

\begin{figure}
    \centering
    \includegraphics[width=1\linewidth]{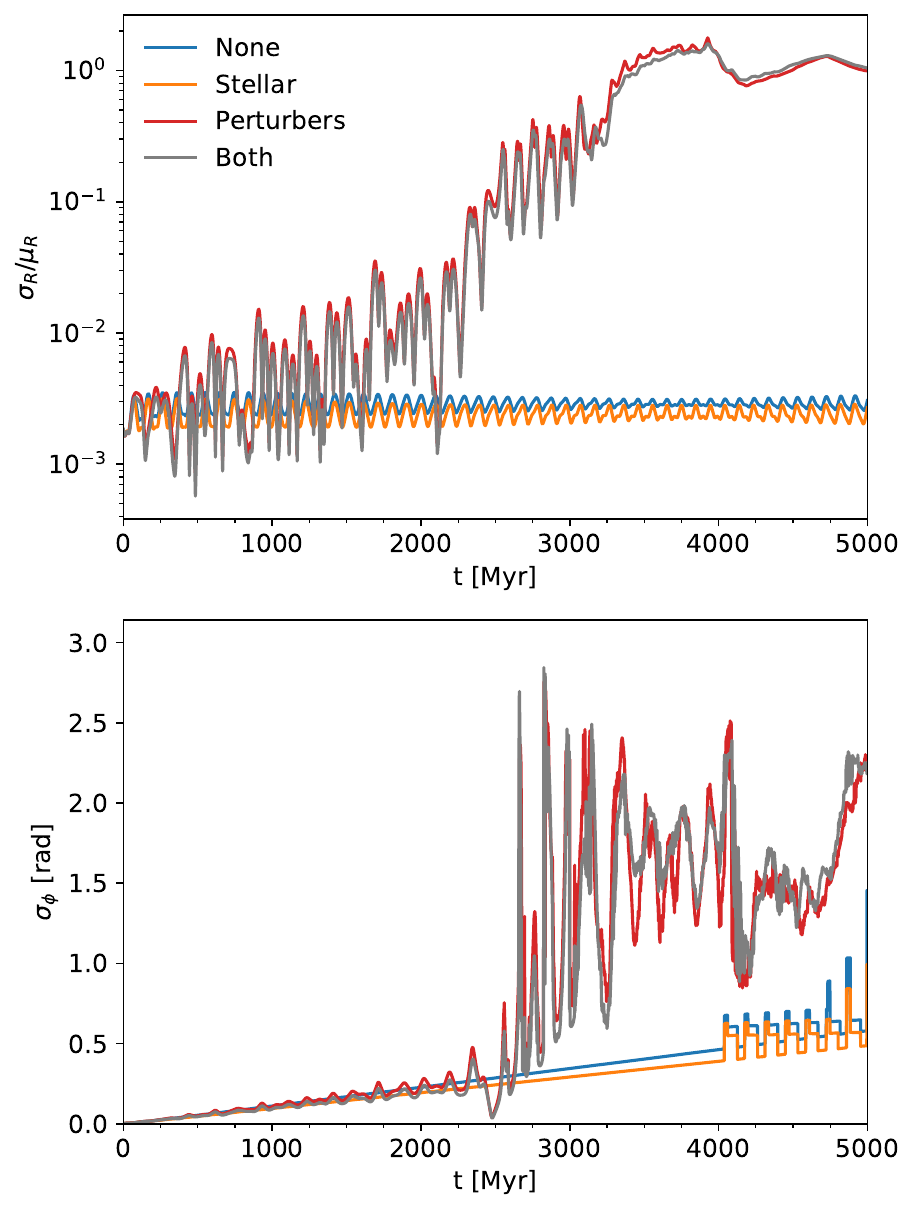}
    \caption{The standard deviation of the distance to the galactic centre normalised to the mean distance (top) and of the azimuthal angle in the galactic plane (bottom) of the cluster members as a function of time. Colour indicates the model setup (red and grey are with perturbers, orange and grey are with stellar evolution).}
    \label{fig:combined_std}
\end{figure}

In Figure \ref{fig:combined_std}, we show the evolution of the star cluster by following the standard deviation of the cluster members' polar coordinates through time. 

If there are no perturbers, the radial standard deviation oscillates but remains below 1\%, meaning that the stars remain close to their initial 5 kpc circular orbit. The azimuthal standard deviation increases mostly linearly, meaning that the cluster steadily disperses along its orbit. After 4 Gyr there are steps in this standard deviation. This is likely a result from phase wrapping; if a considerable fraction of stars has angular coordinates around $-\pi$ or $\pi$, the standard deviation may be biased. Some sections continue the initial linear trend, so the cluster continues to disperse. Both the radial and azimuthal standard deviations are smaller when stellar evolution is included. This is a counterintuitive result, because stellar evolution results in an increasingly unbound cluster, which should promote dispersal into the galaxy.

When perturbers are added, the radial and azimuthal standard deviations increase and become less regular. After 3 Gyr, the radial standard deviation becomes of order unity, implying that the cluster is entirely scattered throughout the galaxy. The azimuthal standard deviation begins to deviate from the case without perturbers at about 2.5 Gyr. Note that the standard deviation of a uniform distribution between $\left[0,2\pi\right]$ is $\frac{\pi}{\sqrt{3}}\approx 1.81$, comparable to the value after 2.5 Gyr.

\begin{figure}
    \centering
    \includegraphics[width=1\linewidth]{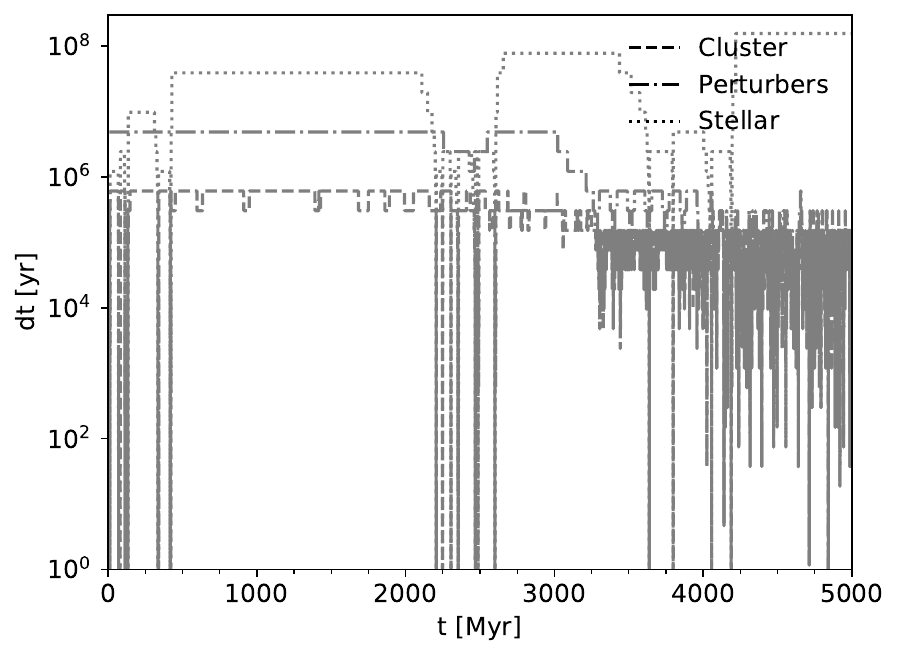}
    \caption{The evolution timescales of different components of the combined model (including both stellar evolution and perturbers) through time. The dashed line indicates the cluster gravity, the dash-dotted line the perturber gravity, and the dotted line stellar evolution.}
    \label{fig:combined_timescales}
\end{figure}

In Figure \ref{fig:combined_timescales}, we show the evolution time-steps of cluster gravity, perturber gravity, and stellar evolution codes through time, for the run with both perturbers and stellar evolution. The stellar time-steps drop below 1 yr during phases of rapid stellar evolution (as seen before in Fig. \ref{fig:coupling_timescale_evolution}), but can also be as large as $\sim$100 Myr. The perturber and cluster time-steps are mostly constant before 3 Gyr (except when the stellar time-steps are small), reflecting that they mostly interact with the galaxy rather than each other. After 3 Gyr, these time-steps become much shorter, and this coincides with the moment the cluster's radial standard deviation reaches unity. This confirms that the cluster's disruption was due to interactions with a perturber.

\section{Discussion}

\subsection{Future applications of the {\sc amuse} implementation}

Although almost any {\sc amuse} simulation could be refactored using
{\sc Venice}, this does not necessarily yield a more efficient code or
a clearer script. {\sc Venice} is especially useful in cases with many
models or a wide (and dynamic) range of timescales. The adaptive
coupling between gravity and stellar evolution could be worked out
further, for example by finding the ideal distribution of stars among
multiple stellar evolution codes depending on their lifetime
distribution. This method could also be used for simulating the
dynamics of smaller systems including planetary systems, such as the
consequences of a planetary system once the parent star turns into a
compact object, or the evolution of massive stellar triples (and
higher-order multiples) in stellar clusters.

{\sc Venice}'s modular expandable nature can be used to gradually and hierarchically build up complex models, allowing both modules and interactions between modules to be characterised individually. A simple planet formation model where a single planet migrates and accretes from a static (gas, pebble, or planetesimal) disk can slowly be expanded to resemble e.g. the Bern model \citep[see Figure 2 in][which has 14 physical modules]{Emsenhuber2021}, where the effect of each module can be characterised individually. Codes can also be exchanged for more efficient or advanced alternatives, which should not affect interactions between other parts of the model that do not interact with that code.

\subsection{Steps towards a differential equation solver}

We introduce {\sc Venice}; a coupling paradigm between numerical
models. The designed method is also suitable for being applied to
differential equation solvers. The main components that have not been
developed yet are the integration of the individual operators
(corresponding with the {\sc evolve\_model} functions) and the
calculation of the coupling timescales $\tau_{ij}$. Here we sketch how
such a method could be constructed.

The individual integration method could be practically any method with time-step control (e.g. Euler (forward or backward), (embedded) Runge-Kutta, or Bulirsch-Stoer), although a single method must be chosen for all operators. The coupling timescale of an operator pair will be the next time-step (predicted by the chosen method) for the differential equation where the right-hand side consists of just those two operators. If these operators are not coupled, this predicted timescale will be determined by the slowest of the two operators. This does not reflect their actual coupling timescale, which is still infinite. This was not an issue in the {\sc amuse} implementation, where the internal evolution time steps were decoupled from the coupling timescale.

The {\sc Venice} algorithm does not eliminate stiff parts of a set of differential equations, but decouples them from non-stiff parts. The integration method should therefore still be implicit in order to maintain stability. A (Bulirsch-Stoer-like) extrapolation method may be useful, because it may allow the slower operator of a fast pair to converge with few iterations, whereas an Euler/Runge-Kutta method will force the slow operator to take the same short time-steps as the fast operator.

A particular application for a {\sc Venice}-based differential equation solver would be a time-dependent chemical network integrator. This consists of many operators (chemical reactions) that operate on a range of timescales depending on reaction rates and chemical abundances.

\section{Conclusion}

We introduced {\sc Venice}, an operator splitting algorithm for
multi-scale, multi-physics numerical problems, and demonstrated its
working with examples from astrophysics. The simple examples show
convergence with decreasing coupling times, although even with
$2^\mathrm{nd}$ order (Strang) coupling this convergence is
potentially only $1^\mathrm{st}$ order, although $2^\mathrm{nd}$ order
coupling typically results in smaller errors (even at a given
wall-clock run-time). We have also shown that updating coupling
timescales can result in appropriate adaptive time-stepping, and that
gravitational {\sc bridge} interactions can be included as a coupling
method. We then demonstrated how a complex model could be built up
with {\sc Venice}.

The {\sc amuse} implementation of {\sc Venice} can now be used as a framework for simulations coupling a large number of numerical models on a wide range of timescales. We have also illustrated how {\sc Venice} can be used to construct a differential equation solver. This requires additional development, such as how to treat the timescales of individual evolution and coupling, and which integration method to use.

\section*{Acknowledgments}

We thank Lourens Veen for the useful discussion on the differences and
similarities between {\sc muscle3} and {\sc venice}, and Shuo Huang
for early user feedback.  MW is supported by NOVA under project number
10.2.5.12.

The simulations in this work have been carried out on a work station
equipped with an Intel Xeon W-2133 CPU, which consumes 140 W. At most,
one out of six cores was used per simulation, putting power usage at
23 W. The amount of computer time used to develop and test {\sc
  Venice} and to produce the figures in this work were not recorded,
but will be on the order of $\sim$300 CPU hours. This puts the total
power consumption at $\sim$6 kWh. At 0.649 kWh/kg of CO$_2$ emitted
(Dutch norm for grey electricity), this comes to $\sim$10 kg of
CO$_2$, comparable to a week of daily commutes.

\bibliographystyle{elsarticle-harv}
\bibliography{venice}

\begin{thebibliography}{39}
\expandafter\ifx\csname natexlab\endcsname\relax\def\natexlab#1{#1}\fi
\providecommand{\url}[1]{\texttt{#1}}
\providecommand{\href}[2]{#2}
\providecommand{\path}[1]{#1}
\providecommand{\DOIprefix}{doi:}
\providecommand{\ArXivprefix}{arXiv:}
\providecommand{\URLprefix}{URL: }
\providecommand{\Pubmedprefix}{pmid:}
\providecommand{\doi}[1]{\href{http://dx.doi.org/#1}{\path{#1}}}
\providecommand{\Pubmed}[1]{\href{pmid:#1}{\path{#1}}}
\providecommand{\bibinfo}[2]{#2}
\ifx\xfnm\relax \def\xfnm[#1]{\unskip,\space#1}\fi
\bibitem[{{Aarseth}(1985)}]{1985mts..conf..377A}
\bibinfo{author}{{Aarseth}, S.J.}, \bibinfo{year}{1985}.
\newblock \bibinfo{title}{{Direct methods for N-body simulations.}}, in:
  \bibinfo{booktitle}{Multiple time scales, p. 377 - 418}, pp.
  \bibinfo{pages}{377--418}.
\bibitem[{Allen et~al.(2001)Allen, Benger, Dramlitsch, Goodale, Hege,
  Lanfermann, Merzky, Radke and Seidel}]{allen2001cactus}
\bibinfo{author}{Allen, G.}, \bibinfo{author}{Benger, W.},
  \bibinfo{author}{Dramlitsch, T.}, \bibinfo{author}{Goodale, T.},
  \bibinfo{author}{Hege, H.C.}, \bibinfo{author}{Lanfermann, G.},
  \bibinfo{author}{Merzky, A.}, \bibinfo{author}{Radke, T.},
  \bibinfo{author}{Seidel, E.}, \bibinfo{year}{2001}.
\newblock \bibinfo{title}{Cactus grid computing: Review of current
  development}, in: \bibinfo{booktitle}{Euro-Par 2001 Parallel Processing: 7th
  International Euro-Par Conference Manchester, UK, August 28--31, 2001
  Proceedings 7}, \bibinfo{organization}{Springer}. pp.
  \bibinfo{pages}{817--824}.
\bibitem[{Babur et~al.(2015)Babur, Smilauer, Verhoeff and van~den
  Brand}]{2015MultiscaleframeworksB}
\bibinfo{author}{Babur, O.}, \bibinfo{author}{Smilauer, V.},
  \bibinfo{author}{Verhoeff, T.}, \bibinfo{author}{van~den Brand, M.},
  \bibinfo{year}{2015}.
\newblock \bibinfo{title}{A survey of open source multiphysics frameworks in
  engineering}, in: \bibinfo{booktitle}{Procedia Computer ScienceVolume 51Issue
  C}, \bibinfo{publisher}{Technische Universiteit Eindhoven}. p.
  \bibinfo{pages}{1088–1097}.
\newblock \DOIprefix\doi{https://doi.org/10.1016/j.procs.2015.05.273}.
\bibitem[{Borgdorff et~al.(2013)Borgdorff, Falcone, Lorenz, Bona-Casas, Chopard
  and Hoekstra}]{Borgdorff2013}
\bibinfo{author}{Borgdorff, J.}, \bibinfo{author}{Falcone, J.L.},
  \bibinfo{author}{Lorenz, E.}, \bibinfo{author}{Bona-Casas, C.},
  \bibinfo{author}{Chopard, B.}, \bibinfo{author}{Hoekstra, A.G.},
  \bibinfo{year}{2013}.
\newblock \bibinfo{title}{Foundations of distributed multiscale computing:
  Formalization, specification, and analysis}.
\newblock \bibinfo{journal}{Journal of Parallel and Distributed Computing}
  \bibinfo{volume}{73}, \bibinfo{pages}{465--483}.
\newblock \URLprefix
  \url{https://www.sciencedirect.com/science/article/pii/S0743731512002924},
  \DOIprefix\doi{https://doi.org/10.1016/j.jpdc.2012.12.011}.
\bibitem[{{Bovy}(2015)}]{Bovy2015}
\bibinfo{author}{{Bovy}, J.}, \bibinfo{year}{2015}.
\newblock \bibinfo{title}{{galpy: A python Library for Galactic Dynamics}}.
\newblock \bibinfo{journal}{\apjs} \bibinfo{volume}{216}, \bibinfo{pages}{29}.
\newblock \DOIprefix\doi{10.1088/0067-0049/216/2/29},
  \href{http://arxiv.org/abs/1412.3451}{{\tt arXiv:1412.3451}}.
\bibitem[{Chopard et~al.(2014)Chopard, Borgdorff and Hoekstra}]{Chopard2014}
\bibinfo{author}{Chopard, B.}, \bibinfo{author}{Borgdorff, J.},
  \bibinfo{author}{Hoekstra, A.}, \bibinfo{year}{2014}.
\newblock \bibinfo{title}{A framework for multi-scale modelling}.
\newblock \bibinfo{journal}{Philosophical transactions. Series A, Mathematical,
  physical, and engineering sciences} \bibinfo{volume}{372}.
\newblock \DOIprefix\doi{10.1098/rsta.2013.0378}.
\bibitem[{Christlieb et~al.(2015)Christlieb, Liu and Xu}]{Christlieb2015}
\bibinfo{author}{Christlieb, A.J.}, \bibinfo{author}{Liu, Y.},
  \bibinfo{author}{Xu, Z.}, \bibinfo{year}{2015}.
\newblock \bibinfo{title}{High order operator splitting methods based on an
  integral deferred correction framework}.
\newblock \bibinfo{journal}{Journal of Computational Physics}
  \bibinfo{volume}{294}, \bibinfo{pages}{224--242}.
\newblock \URLprefix
  \url{https://www.sciencedirect.com/science/article/pii/S0021999115001795},
  \DOIprefix\doi{https://doi.org/10.1016/j.jcp.2015.03.032}.
\bibitem[{Einarsrud et~al.(2017)Einarsrud, Eick, Bai, Feng, Hua and
  Witt}]{EINARSRUD20173}
\bibinfo{author}{Einarsrud, K.E.}, \bibinfo{author}{Eick, I.},
  \bibinfo{author}{Bai, W.}, \bibinfo{author}{Feng, Y.}, \bibinfo{author}{Hua,
  J.}, \bibinfo{author}{Witt, P.J.}, \bibinfo{year}{2017}.
\newblock \bibinfo{title}{Towards a coupled multi-scale, multi-physics
  simulation framework for aluminium electrolysis}.
\newblock \bibinfo{journal}{Applied Mathematical Modelling}
  \bibinfo{volume}{44}, \bibinfo{pages}{3--24}.
\newblock \URLprefix
  \url{https://www.sciencedirect.com/science/article/pii/S0307904X16306060},
  \DOIprefix\doi{https://doi.org/10.1016/j.apm.2016.11.011}.
\bibitem[{{Emsenhuber} et~al.(2021){Emsenhuber}, {Mordasini}, {Burn},
  {Alibert}, {Benz} and {Asphaug}}]{Emsenhuber2021}
\bibinfo{author}{{Emsenhuber}, A.}, \bibinfo{author}{{Mordasini}, C.},
  \bibinfo{author}{{Burn}, R.}, \bibinfo{author}{{Alibert}, Y.},
  \bibinfo{author}{{Benz}, W.}, \bibinfo{author}{{Asphaug}, E.},
  \bibinfo{year}{2021}.
\newblock \bibinfo{title}{{The New Generation Planetary Population Synthesis
  (NGPPS). I. Bern global model of planet formation and evolution, model tests,
  and emerging planetary systems}}.
\newblock \bibinfo{journal}{\aap} \bibinfo{volume}{656}, \bibinfo{pages}{A69}.
\newblock \DOIprefix\doi{10.1051/0004-6361/202038553},
  \href{http://arxiv.org/abs/2007.05561}{{\tt arXiv:2007.05561}}.
\bibitem[{{Fujii} et~al.(2007){Fujii}, {Iwasawa}, {Funato} and
  {Makino}}]{Fujii2007}
\bibinfo{author}{{Fujii}, M.}, \bibinfo{author}{{Iwasawa}, M.},
  \bibinfo{author}{{Funato}, Y.}, \bibinfo{author}{{Makino}, J.},
  \bibinfo{year}{2007}.
\newblock \bibinfo{title}{{BRIDGE: A Direct-Tree Hybrid N-Body Algorithm for
  Fully Self-Consistent Simulations of Star Clusters and Their Parent
  Galaxies}}.
\newblock \bibinfo{journal}{\pasj} \bibinfo{volume}{59}, \bibinfo{pages}{1095}.
\newblock \DOIprefix\doi{10.1093/pasj/59.6.109510.48550/arXiv.0706.2059},
  \href{http://arxiv.org/abs/0706.2059}{{\tt arXiv:0706.2059}}.
\bibitem[{Goodale et~al.(2003)Goodale, Allen, Lanfermann, Mass{\'o}, Radke,
  Seidel and Shalf}]{10.1007/3-540-36569-9_13}
\bibinfo{author}{Goodale, T.}, \bibinfo{author}{Allen, G.},
  \bibinfo{author}{Lanfermann, G.}, \bibinfo{author}{Mass{\'o}, J.},
  \bibinfo{author}{Radke, T.}, \bibinfo{author}{Seidel, E.},
  \bibinfo{author}{Shalf, J.}, \bibinfo{year}{2003}.
\newblock \bibinfo{title}{The cactus framework and toolkit: Design and
  applications}, in: \bibinfo{editor}{Palma, J.M.L.M.}, \bibinfo{editor}{Sousa,
  A.A.}, \bibinfo{editor}{Dongarra, J.}, \bibinfo{editor}{Hern{\'a}ndez, V.}
  (Eds.), \bibinfo{booktitle}{High Performance Computing for Computational
  Science --- VECPAR 2002}, \bibinfo{publisher}{Springer Berlin Heidelberg},
  \bibinfo{address}{Berlin, Heidelberg}. pp. \bibinfo{pages}{197--227}.
\bibitem[{{Huang} et~al.(2024){Huang}, Portegies~Zwart and
  Wilhelm}]{Shuo2etal2024}
\bibinfo{author}{{Huang}, S.}, \bibinfo{author}{Portegies~Zwart, S.},
  \bibinfo{author}{Wilhelm, M.}, \bibinfo{year}{2024}.
\newblock \bibinfo{title}{{The formation and evolution of planetary systems in
  dense stallar clusters}}.
\newblock \bibinfo{journal}{submitted to A\&A} .
\bibitem[{{J{\"a}nes} et~al.(2014){J{\"a}nes}, {Pelupessy} and {Portegies
  Zwart}}]{Janes2014}
\bibinfo{author}{{J{\"a}nes}, J.}, \bibinfo{author}{{Pelupessy}, I.},
  \bibinfo{author}{{Portegies Zwart}, S.}, \bibinfo{year}{2014}.
\newblock \bibinfo{title}{{A connected component-based method for efficiently
  integrating multi-scale N-body systems}}.
\newblock \bibinfo{journal}{\aap} \bibinfo{volume}{570}, \bibinfo{pages}{A20}.
\newblock \DOIprefix\doi{10.1051/0004-6361/20142383110.48550/arXiv.1407.7105},
  \href{http://arxiv.org/abs/1407.7105}{{\tt arXiv:1407.7105}}.
\bibitem[{Jia and Li(2011)}]{Jia2011}
\bibinfo{author}{Jia, H.}, \bibinfo{author}{Li, K.}, \bibinfo{year}{2011}.
\newblock \bibinfo{title}{A third accurate operator splitting method}.
\newblock \bibinfo{journal}{Mathematical and Computer Modelling}
  \bibinfo{volume}{53}, \bibinfo{pages}{387--396}.
\newblock \URLprefix
  \url{https://www.sciencedirect.com/science/article/pii/S089571771000436X},
  \DOIprefix\doi{https://doi.org/10.1016/j.mcm.2010.09.005}.
\bibitem[{{Kroupa}(2001)}]{Kroupa2001}
\bibinfo{author}{{Kroupa}, P.}, \bibinfo{year}{2001}.
\newblock \bibinfo{title}{{On the variation of the initial mass function}}.
\newblock \bibinfo{journal}{\mnras} \bibinfo{volume}{322},
  \bibinfo{pages}{231--246}.
\newblock \DOIprefix\doi{10.1046/j.1365-8711.2001.04022.x},
  \href{http://arxiv.org/abs/astro-ph/0009005}{{\tt arXiv:astro-ph/0009005}}.
\bibitem[{{McMillan} et~al.(2012){McMillan}, {Portegies Zwart}, {van Elteren}
  and {Whitehead}}]{McMillan2012}
\bibinfo{author}{{McMillan}, S.}, \bibinfo{author}{{Portegies Zwart}, S.},
  \bibinfo{author}{{van Elteren}, A.}, \bibinfo{author}{{Whitehead}, A.},
  \bibinfo{year}{2012}.
\newblock \bibinfo{title}{{Simulations of Dense Stellar Systems with the AMUSE
  Software Toolkit}}, in: \bibinfo{editor}{{Capuzzo-Dolcetta}, R.},
  \bibinfo{editor}{{Limongi}, M.}, \bibinfo{editor}{{Tornamb{\`e}}, A.} (Eds.),
  \bibinfo{booktitle}{Advances in Computational Astrophysics: Methods, Tools,
  and Outcome}, p. \bibinfo{pages}{129}.
\newblock \href{http://arxiv.org/abs/1111.3987}{{\tt arXiv:1111.3987}}.
\bibitem[{{McMillan}(1986)}]{McMillan1986}
\bibinfo{author}{{McMillan}, S.L.W.}, \bibinfo{year}{1986}.
\newblock \bibinfo{title}{{The Vectorization of Small-N Integrators}}, in:
  \bibinfo{editor}{{Hut}, P.}, \bibinfo{editor}{{McMillan}, S.L.W.} (Eds.),
  \bibinfo{booktitle}{The Use of Supercomputers in Stellar Dynamics}. volume
  \bibinfo{volume}{267}, p. \bibinfo{pages}{156}.
\newblock \DOIprefix\doi{10.1007/BFb0116406}.
\bibitem[{{Miller}(1964)}]{Miller1964}
\bibinfo{author}{{Miller}, R.H.}, \bibinfo{year}{1964}.
\newblock \bibinfo{title}{{Irreversibility in Small Stellar Dynamical
  Systems.}}
\newblock \bibinfo{journal}{\apj} \bibinfo{volume}{140}, \bibinfo{pages}{250}.
\newblock \DOIprefix\doi{10.1086/147911}.
\bibitem[{{Miyamoto} and {Nagai}(1975)}]{Miyamoto1975}
\bibinfo{author}{{Miyamoto}, M.}, \bibinfo{author}{{Nagai}, R.},
  \bibinfo{year}{1975}.
\newblock \bibinfo{title}{{Three-dimensional models for the distribution of
  mass in galaxies.}}
\newblock \bibinfo{journal}{\pasj} \bibinfo{volume}{27},
  \bibinfo{pages}{533--543}.
\bibitem[{{Navarro} et~al.(1996){Navarro}, {Frenk} and {White}}]{Navarro1996}
\bibinfo{author}{{Navarro}, J.F.}, \bibinfo{author}{{Frenk}, C.S.},
  \bibinfo{author}{{White}, S.D.M.}, \bibinfo{year}{1996}.
\newblock \bibinfo{title}{{The Structure of Cold Dark Matter Halos}}.
\newblock \bibinfo{journal}{\apj} \bibinfo{volume}{462}, \bibinfo{pages}{563}.
\newblock \DOIprefix\doi{10.1086/177173},
  \href{http://arxiv.org/abs/astro-ph/9508025}{{\tt arXiv:astro-ph/9508025}}.
\bibitem[{{Pelupessy} et~al.(2012){Pelupessy}, {J{\"a}nes} and {Portegies
  Zwart}}]{Pelupessy2012a}
\bibinfo{author}{{Pelupessy}, F.I.}, \bibinfo{author}{{J{\"a}nes}, J.},
  \bibinfo{author}{{Portegies Zwart}, S.}, \bibinfo{year}{2012}.
\newblock \bibinfo{title}{{N-body integrators with individual time steps from
  Hierarchical splitting}}.
\newblock \bibinfo{journal}{\na} \bibinfo{volume}{17},
  \bibinfo{pages}{711--719}.
\newblock \DOIprefix\doi{10.1016/j.newast.2012.05.009},
  \href{http://arxiv.org/abs/1205.5668}{{\tt arXiv:1205.5668}}.
\bibitem[{{Pelupessy} and {Portegies Zwart}(2012)}]{Pelupessy2012b}
\bibinfo{author}{{Pelupessy}, F.I.}, \bibinfo{author}{{Portegies Zwart}, S.},
  \bibinfo{year}{2012}.
\newblock \bibinfo{title}{{The evolution of embedded star clusters}}.
\newblock \bibinfo{journal}{\mnras} \bibinfo{volume}{420},
  \bibinfo{pages}{1503--1517}.
\newblock \DOIprefix\doi{10.1111/j.1365-2966.2011.20137.x},
  \href{http://arxiv.org/abs/1111.0992}{{\tt arXiv:1111.0992}}.
\bibitem[{{Pelupessy} et~al.(2013){Pelupessy}, {van Elteren}, {de Vries},
  {McMillan}, {Drost} and {Portegies Zwart}}]{Pelupessy2013}
\bibinfo{author}{{Pelupessy}, F.I.}, \bibinfo{author}{{van Elteren}, A.},
  \bibinfo{author}{{de Vries}, N.}, \bibinfo{author}{{McMillan}, S.L.W.},
  \bibinfo{author}{{Drost}, N.}, \bibinfo{author}{{Portegies Zwart}, S.F.},
  \bibinfo{year}{2013}.
\newblock \bibinfo{title}{{The Astrophysical Multipurpose Software
  Environment}}.
\newblock \bibinfo{journal}{\aap} \bibinfo{volume}{557}, \bibinfo{pages}{A84}.
\newblock \DOIprefix\doi{10.1051/0004-6361/201321252},
  \href{http://arxiv.org/abs/1307.3016}{{\tt arXiv:1307.3016}}.
\bibitem[{{Peters} et~al.(2018){Peters}, {Baniasadi}, {Baniasadi}, {Besseron},
  {Estupinan Donoso}, {Mohseni} and {Pozzetti}}]{2018arXiv180808028P}
\bibinfo{author}{{Peters}, B.}, \bibinfo{author}{{Baniasadi}, M.},
  \bibinfo{author}{{Baniasadi}, M.}, \bibinfo{author}{{Besseron}, X.},
  \bibinfo{author}{{Estupinan Donoso}, A.}, \bibinfo{author}{{Mohseni}, M.},
  \bibinfo{author}{{Pozzetti}, G.}, \bibinfo{year}{2018}.
\newblock \bibinfo{title}{{The XDEM Multi-physics and Multi-scale Simulation
  Technology: Review on DEM-CFD Coupling, Methodology and Engineering
  Applications}}.
\newblock \bibinfo{journal}{arXiv e-prints} ,
  \bibinfo{pages}{arXiv:1808.08028}\DOIprefix\doi{10.48550/arXiv.1808.08028},
  \href{http://arxiv.org/abs/1808.08028}{{\tt arXiv:1808.08028}}.
\bibitem[{{Portegies Zwart}(2018)}]{PortegiesZwart2018}
\bibinfo{author}{{Portegies Zwart}, S.}, \bibinfo{year}{2018}.
\newblock \bibinfo{title}{{Computational astrophysics for the future}}.
\newblock \bibinfo{journal}{Science} \bibinfo{volume}{361},
  \bibinfo{pages}{979--980}.
\newblock \DOIprefix\doi{10.1126/science.aau3206},
  \href{http://arxiv.org/abs/1809.02600}{{\tt arXiv:1809.02600}}.
\bibitem[{{Portegies Zwart} et~al.(2009){Portegies Zwart}, {McMillan},
  {Harfst}, {Groen}, {Fujii}, {Nuall{\'a}in}, {Glebbeek}, {Heggie}, {Lombardi},
  {Hut}, {Angelou}, {Banerjee}, {Belkus}, {Fragos}, {Fregeau}, {Gaburov},
  {Izzard}, {Juri{\'c}}, {Justham}, {Sottoriva}, {Teuben}, {van Bever}, {Yaron}
  and {Zemp}}]{PortegiesZwart2009}
\bibinfo{author}{{Portegies Zwart}, S.}, \bibinfo{author}{{McMillan}, S.},
  \bibinfo{author}{{Harfst}, S.}, \bibinfo{author}{{Groen}, D.},
  \bibinfo{author}{{Fujii}, M.}, \bibinfo{author}{{Nuall{\'a}in}, B.{\'O}.},
  \bibinfo{author}{{Glebbeek}, E.}, \bibinfo{author}{{Heggie}, D.},
  \bibinfo{author}{{Lombardi}, J.}, \bibinfo{author}{{Hut}, P.},
  \bibinfo{author}{{Angelou}, V.}, \bibinfo{author}{{Banerjee}, S.},
  \bibinfo{author}{{Belkus}, H.}, \bibinfo{author}{{Fragos}, T.},
  \bibinfo{author}{{Fregeau}, J.}, \bibinfo{author}{{Gaburov}, E.},
  \bibinfo{author}{{Izzard}, R.}, \bibinfo{author}{{Juri{\'c}}, M.},
  \bibinfo{author}{{Justham}, S.}, \bibinfo{author}{{Sottoriva}, A.},
  \bibinfo{author}{{Teuben}, P.}, \bibinfo{author}{{van Bever}, J.},
  \bibinfo{author}{{Yaron}, O.}, \bibinfo{author}{{Zemp}, M.},
  \bibinfo{year}{2009}.
\newblock \bibinfo{title}{{A multiphysics and multiscale software environment
  for modeling astrophysical systems}}.
\newblock \bibinfo{journal}{\na} \bibinfo{volume}{14},
  \bibinfo{pages}{369--378}.
\newblock \DOIprefix\doi{10.1016/j.newast.2008.10.006},
  \href{http://arxiv.org/abs/0807.1996}{{\tt arXiv:0807.1996}}.
\bibitem[{{Portegies Zwart} et~al.(2013){Portegies Zwart}, {McMillan}, {van
  Elteren}, {Pelupessy} and {de Vries}}]{PortegiesZwart2013}
\bibinfo{author}{{Portegies Zwart}, S.}, \bibinfo{author}{{McMillan}, S.L.W.},
  \bibinfo{author}{{van Elteren}, E.}, \bibinfo{author}{{Pelupessy}, I.},
  \bibinfo{author}{{de Vries}, N.}, \bibinfo{year}{2013}.
\newblock \bibinfo{title}{{Multi-physics simulations using a hierarchical
  interchangeable software interface}}.
\newblock \bibinfo{journal}{Computer Physics Communications}
  \bibinfo{volume}{184}, \bibinfo{pages}{456--468}.
\newblock \DOIprefix\doi{10.1016/j.cpc.2012.09.024},
  \href{http://arxiv.org/abs/1204.5522}{{\tt arXiv:1204.5522}}.
\bibitem[{{Portegies Zwart} et~al.(2020a){Portegies Zwart}, {Pelupessy},
  {Mart{\'\i}nez-Barbosa}, {van Elteren} and {McMillan}}]{PortegiesZwart2020}
\bibinfo{author}{{Portegies Zwart}, S.}, \bibinfo{author}{{Pelupessy}, I.},
  \bibinfo{author}{{Mart{\'\i}nez-Barbosa}, C.}, \bibinfo{author}{{van
  Elteren}, A.}, \bibinfo{author}{{McMillan}, S.}, \bibinfo{year}{2020}a.
\newblock \bibinfo{title}{{Non-intrusive hierarchical coupling strategies for
  multi-scale simulations in gravitational dynamics}}.
\newblock \bibinfo{journal}{Communications in Nonlinear Science and Numerical
  Simulations} \bibinfo{volume}{85}, \bibinfo{pages}{105240}.
\newblock \DOIprefix\doi{10.1016/j.cnsns.2020.105240},
  \href{http://arxiv.org/abs/2002.11206}{{\tt arXiv:2002.11206}}.
\bibitem[{{Portegies Zwart} et~al.(2020b){Portegies Zwart}, {Pelupessy},
  {Mart{\'\i}nez-Barbosa}, {van Elteren} and {McMillan}}]{2020CNSNS..8505240P}
\bibinfo{author}{{Portegies Zwart}, S.}, \bibinfo{author}{{Pelupessy}, I.},
  \bibinfo{author}{{Mart{\'\i}nez-Barbosa}, C.}, \bibinfo{author}{{van
  Elteren}, A.}, \bibinfo{author}{{McMillan}, S.}, \bibinfo{year}{2020}b.
\newblock \bibinfo{title}{{Non-intrusive hierarchical coupling strategies for
  multi-scale simulations in gravitational dynamics}}.
\newblock \bibinfo{journal}{Communications in Nonlinear Science and Numerical
  Simulations} \bibinfo{volume}{85}, \bibinfo{pages}{105240}.
\newblock \DOIprefix\doi{10.1016/j.cnsns.2020.105240},
  \href{http://arxiv.org/abs/2002.11206}{{\tt arXiv:2002.11206}}.
\bibitem[{{Portegies Zwart} and {Verbunt}(1996)}]{PortegiesZwart1996}
\bibinfo{author}{{Portegies Zwart}, S.F.}, \bibinfo{author}{{Verbunt}, F.},
  \bibinfo{year}{1996}.
\newblock \bibinfo{title}{{Population synthesis of high-mass binaries.}}
\newblock \bibinfo{journal}{\aap} \bibinfo{volume}{309},
  \bibinfo{pages}{179--196}.
\bibitem[{Rin et~al.(2017)Rin, Tomin, Garipov, Voskov and
  Tchelepi}]{10.2118/182714-MS}
\bibinfo{editor}{Rin, R.}, \bibinfo{editor}{Tomin, P.},
  \bibinfo{editor}{Garipov, T.}, \bibinfo{editor}{Voskov, D.},
  \bibinfo{editor}{Tchelepi, H.} (Eds.), \bibinfo{year}{2017}.
\newblock \bibinfo{title}{{General Implicit Coupling Framework for
  Multi-Physics Problems}}. volume \bibinfo{volume}{Day 3 Wed, February 22,
  2017} of \textit{\bibinfo{series}{SPE Reservoir Simulation Conference}}.
\newblock \URLprefix \url{https://doi.org/10.2118/182714-MS},
  \DOIprefix\doi{10.2118/182714-MS},
  \href{http://arxiv.org/abs/https://onepetro.org/spersc/proceedings-pdf/17RSC/3-17RSC/D031S012R001/1298640/spe-182714-ms.pdf}{{\tt
  arXiv:https://onepetro.org/spersc/proceedings-pdf/17RSC/3-17RSC/D031S012R001/1298640/spe-182714-ms.pdf}}.
\bibitem[{{Strang}(1968)}]{Strang1968}
\bibinfo{author}{{Strang}, G.}, \bibinfo{year}{1968}.
\newblock \bibinfo{title}{{On the Construction and Comparison of Difference
  Schemes}}.
\newblock \bibinfo{journal}{SIAM Journal on Numerical Analysis}
  \bibinfo{volume}{5}, \bibinfo{pages}{506--517}.
\newblock \DOIprefix\doi{10.1137/0705041}.
\bibitem[{{Terlevich}(1987)}]{Terlevich1987}
\bibinfo{author}{{Terlevich}, E.}, \bibinfo{year}{1987}.
\newblock \bibinfo{title}{{Evolution of N-body open clusters.}}
\newblock \bibinfo{journal}{\mnras} \bibinfo{volume}{224},
  \bibinfo{pages}{193--225}.
\newblock \DOIprefix\doi{10.1093/mnras/224.1.193}.
\bibitem[{{Toonen} et~al.(2012){Toonen}, {Nelemans} and {Portegies
  Zwart}}]{Toonen2012}
\bibinfo{author}{{Toonen}, S.}, \bibinfo{author}{{Nelemans}, G.},
  \bibinfo{author}{{Portegies Zwart}, S.}, \bibinfo{year}{2012}.
\newblock \bibinfo{title}{{Supernova Type Ia progenitors from merging double
  white dwarfs. Using a new population synthesis model}}.
\newblock \bibinfo{journal}{\aap} \bibinfo{volume}{546}, \bibinfo{pages}{A70}.
\newblock \DOIprefix\doi{10.1051/0004-6361/201218966},
  \href{http://arxiv.org/abs/1208.6446}{{\tt arXiv:1208.6446}}.
\bibitem[{{van Elteren} et~al.(2019){van Elteren}, {Portegies Zwart},
  {Pelupessy}, {Cai} and {McMillan}}]{2019A&A...624A.120V}
\bibinfo{author}{{van Elteren}, A.}, \bibinfo{author}{{Portegies Zwart}, S.},
  \bibinfo{author}{{Pelupessy}, I.}, \bibinfo{author}{{Cai}, M.X.},
  \bibinfo{author}{{McMillan}, S.L.W.}, \bibinfo{year}{2019}.
\newblock \bibinfo{title}{{Survivability of planetary systems in young and
  dense star clusters}}.
\newblock \bibinfo{journal}{\aap} \bibinfo{volume}{624}, \bibinfo{pages}{A120}.
\newblock \DOIprefix\doi{10.1051/0004-6361/201834641},
  \href{http://arxiv.org/abs/1902.04652}{{\tt arXiv:1902.04652}}.
\bibitem[{Veen and Hoekstra(2020)}]{Veen2020}
\bibinfo{author}{Veen, L.E.}, \bibinfo{author}{Hoekstra, A.G.},
  \bibinfo{year}{2020}.
\newblock \bibinfo{title}{Easing multiscale model design and coupling with
  muscle 3}, in: \bibinfo{editor}{Krzhizhanovskaya, V.V.},
  \bibinfo{editor}{Z{\'a}vodszky, G.}, \bibinfo{editor}{Lees, M.H.},
  \bibinfo{editor}{Dongarra, J.J.}, \bibinfo{editor}{Sloot, P.M.A.},
  \bibinfo{editor}{Brissos, S.}, \bibinfo{editor}{Teixeira, J.} (Eds.),
  \bibinfo{booktitle}{Computational Science -- ICCS 2020},
  \bibinfo{publisher}{Springer International Publishing},
  \bibinfo{address}{Cham}. pp. \bibinfo{pages}{425--438}.
\bibitem[{Verlet(1967)}]{PhysRev.159.98}
\bibinfo{author}{Verlet, L.}, \bibinfo{year}{1967}.
\newblock \bibinfo{title}{Computer "experiments" on classical fluids. i.
  thermodynamical properties of lennard-jones molecules}.
\newblock \bibinfo{journal}{Phys. Rev.} \bibinfo{volume}{159},
  \bibinfo{pages}{98--103}.
\newblock \URLprefix \url{http://link.aps.org/doi/10.1103/PhysRev.159.98},
  \DOIprefix\doi{10.1103/PhysRev.159.98}.
\bibitem[{{Wielen}(1967)}]{Wielen1967}
\bibinfo{author}{{Wielen}, R.}, \bibinfo{year}{1967}.
\newblock \bibinfo{title}{{Dynamical Evolution of Star Cluster Models, I.}}
\newblock \bibinfo{journal}{Veroeffentlichungen des Astronomischen
  Rechen-Instituts Heidelberg} \bibinfo{volume}{19}, \bibinfo{pages}{1}.
\bibitem[{Xianmeng et~al.(2021)Xianmeng, Mingyu, Xiao, Zhaoshun, Yinyu, Xu and
  Suxuan}]{10.1177/0037549719881204}
\bibinfo{author}{Xianmeng, W.}, \bibinfo{author}{Mingyu, W.},
  \bibinfo{author}{Xiao, H.}, \bibinfo{author}{Zhaoshun, W.},
  \bibinfo{author}{Yinyu, C.}, \bibinfo{author}{Xu, L.},
  \bibinfo{author}{Suxuan, G.}, \bibinfo{year}{2021}.
\newblock \bibinfo{title}{Multi-physics coupling simulation in virtual
  reactors}.
\newblock \bibinfo{journal}{Simulation} \bibinfo{volume}{97},
  \bibinfo{pages}{687–702}.
\newblock \URLprefix \url{https://doi.org/10.1177/0037549719881204},
  \DOIprefix\doi{10.1177/0037549719881204}.

\end{thebibliography}

\end{document}